\documentclass[a4paper,aip,jcp,preprint,superscriptaddress]{revtex4-1}

\usepackage{amsfonts}

\newcommand{\Td}{\mathrm{T}_d}
\newcommand{\methane}{$\mathrm{XY_4}$}

\begin{document}

\title{A Fast Algorithm for the Construction of Integrity Bases Associated to
  Symmetry--Adapted Polynomial Representations. Application to Tetrahedral
  \methane{} Molecules}

\author{Patrick Cassam-Chena\"\i}
\email{cassam@unice.fr}
\affiliation{Univ. Nice Sophia Antipolis, CNRS,  LJAD, UMR 7351, 06100 Nice,
  France.}
\author{Guillaume Dhont}
%\email{guillaume.dhont@univ-littoral.fr}
\affiliation{Laboratoire de Physico--Chimie de l'Atmosph\`ere,
MREI~2,189A Avenue Maurice Schumann,
59140 Dunkerque,
France.}
\author{Fr\'ed\'eric Patras}
%\email{patras@unice.fr}
\affiliation{Univ. Nice Sophia Antipolis, CNRS,  LJAD, UMR 7351, 06100 Nice,
  France.}
\date{\today}


\begin{abstract}
Invariant theory provides more efficient tools, such as Molien generating
functions and integrity bases, than basic group theory, that relies on
projector techniques for the construction of symmetry--adapted polynomials in
the symmetry coordinates of a molecular system, because it is based on a finer
description of the mathematical structure of the latter.  
The present article extends its use to the construction of polynomial bases
which span possibly, non--totally symmetric irreducible representations of a
molecular symmetry group.  Electric or magnetic observables can carry such
irreducible representations, a common example is given by the electric dipole
moment surface.
% We present a fast
% algorithm to construct the integrity basis for the polynomials that transform
% according to a given non--totally symmetric irreducible representation. 
The elementary generating functions and their corresponding integrity bases,
where both the initial and the final representations are irreducible, are the
building blocks of the algorithm presented in this article, which is faster
than algorithms based on projection operators only.  The generating functions
for the full initial representation of interest are built recursively from the
elementary generating functions.  Integrity bases which can be used to
generate in the most economical way symmetry--adapted polynomial bases are
constructed alongside in the same fashion.
% Our algorithm follows this
% step--by--step building up of the final generating function and constructs
% accordingly the required integrity basis.  
The method is illustrated in detail on \methane{} type of molecules.  Explicit
integrity bases for all five possible final irreducible representations of the
tetrahedral group have been calculated and are given in the supplemental
material associated with this paper.
\end{abstract} 
\maketitle

\section{Introduction}
The simulation of the rotation--vibration molecular spectrum requires the
knowledge of the potential energy surface (PES) and of the electric dipole
moment surface (EDMS) of the molecule under study. These two functions of
internal coordinates do not have a known analytic expression. This issue is
often encountered in quantum chemistry or computational spectroscopy and a
typical solution is to expand these functions on a set of appropriate
analytical functions. The expansion coefficients are then determined
by fitting over experimental or theoretical data. Symmetry
helps to simplify the problem\cite{b101,b169,b163,b019,a0629} and favors the
introduction of symmetry--adapted coordinates when the function to be expanded
transforms according to an irreducible representation (irrep) of the symmetry
group $G$ of the molecule.  In particular, the PES transforms as the totally
symmetric (also called trivial) irrep of the group $G$ while the components of
the EDMS may carry a non--trivial representation of the group.

The set of symmetrized internal coordinates spans a representation called the
initial, usually reducible, representation $\Gamma^\mathrm{initial}$.
Symmetry--adapted polynomials in these variables are then considered.  The
polynomials that transform according to the final irrep
$\Gamma^\mathrm{final}$ are called $\Gamma^\mathrm{final}$--covariant
polynomials.\cite{b100} An ``invariant'' polynomial is the distinct case of this
classification when $\Gamma^\mathrm{final}$ is the totally symmetric
representation of the group, noted $A$, $A^\prime$, $A_1$, or $A_g$ in 
character tables.

Projecting on irreps using projection operators is a standard method of group
theory to generate symmetry--adapted polynomials. Marquardt\cite{doc_03183}
and Schwenke\cite{doc_03720} relied on this technique to compute
symmetry--adapted basis sets and expand the PES of methane.  The projection
method for the construction of invariants is applicable to irreps of dimension
higher than one through the introduction of projection operators together with
transfer operators, see Hamermesh,\cite{b064} Bunker,~\cite{b019}
Lomont,\cite{doc_03732} and Taylor.\cite{doc_03734} The group--theoretical
methods based on projectors are inherently inefficient because they
ignore the number of linearly independent symmetry--adapted polynomials of a
given degree $k$. So, in order to obtain a complete set, they have to consider
all possible starting polynomial ``seeds'', usually a basis set of
monomials. The projection of the latters often lead to the null polynomial or
to a useless linear combination of already known symmetry--adapted
polynomials.  Furthermore, the dimension of the space of symmetry--adapted
polynomials becomes rapidly formidable even at modest $k$ and the list of
polynomials to tabulate becomes unnecessarily gigantic.

Another technique of construction of symmetry--adapted polynomials is based on
the Clebsch--Gordan coefficients of the group $G$.  A great deal of work has
been dedicated in particular to the cubic
group.\cite{doc_03458,doc_03459,doc_03470,doc_03595} The coupling with 
Clebsch--Gordan coefficients of two polynomials give a polynomial of higher
degree and the set of symmetry--adapted polynomials is built degree by
degree. All possible couplings between vector space basis sets of polynomials
of lower degrees must be considered to insure that one gets a complete
list. Compared to the previous approach based on projection operators, the
computational effort is reduced but the tabulated basis sets have the same
unnecessarily large sizes.

The drawbacks of the two approaches described above are circumvented by
polynomial ring invariant theory, which in spite of its name encompasses the
covariant case and fully exploits the algebraic structure of
$\Gamma^\mathrm{final}$--covariant polynomials. In particular, the coefficient
$c_k$ of the Taylor expansion $c_0+c_1t+c_2t^2+c_3t^3+\cdots$ of the Molien
generating function\cite{a0628,a1776} gives the number of linearly independent
polynomials of degree $k$ carrying a given symmetry. The introduction of
invariant theory in quantum chemistry can be traced back to the works of
Murrell \textit{et al.}.\cite{doc_03084,doc_01779} Followers include Collins
and Parsons,\cite{a1610} Ischtwan and Peyerimhoff.\cite{doc_01780} Recently,
Braams, Bowman and their collaborators introduced permutationally invariant
polynomial bases that satisfy the permutational symmetry with respect to
like atoms.  \cite{doc_03613,doc_03158,doc_03622} However, these
studies were only concerned with the totally symmetric representation in
relation to the expansion of a PES. Braams and Bowman did consider expansions
of an EDMS but they reduced the problem to the totally symmetric case by
restricting themselves to a subgroup of the molecular point group, which is
not optimal.

An integrity basis for $\Gamma^\mathrm{final}$--covariant polynomials
involves two finite sets of polynomials.\cite{a0628,doc_03181} The first set
contains $D$ \textit{denominator} or \textit{primary} polynomials $f_i$,
$1\leq i\leq D$, which are algebraically independent invariant
polynomials.\cite{a0628,b123} The second set contains $N$ linearly independent
\textit{numerator} or \textit{secondary} polynomials $g_j$, $1\leq j\leq N$,
which transform as the $\Gamma^\mathrm{final}$ representation. Any
$\Gamma^\mathrm{final}$--covariant polynomial $p$ admits a unique
decomposition in the denominator and numerator polynomials: $p=g_1\times
h_1\left(f_1,\ldots,f_D\right) + \cdots + g_N\times
h_N\left(f_1,\ldots,f_D\right)$.  The $h_j$ are polynomials in $D$ variables:
any nonnegative integer can be a power of the denominator polynomials while
numerator polynomials only appear linearly.  The important result is that the
integrity basis is a much more compact way to present the set of
$\Gamma^\mathrm{final}$--covariant polynomials than a list of vector space
bases for each degree $k$. All the $\Gamma^\mathrm{final}$--covariant
polynomials, up to any order, can be generated from the polynomials belonging
to the small integrity basis by a direct algorithm. This circumvents the
problems inherent to projector or Clebsch--Gordan based methods, where
gigantic tables necessarily limited to a given (usually low) degree have to be
stored.  Applications of integrity bases are numerous. They have been used to
define error--correcting codes in applied mathematics,\cite{a1611} to analyze
problems involving crystal symmetry,\cite{a1462,a1463} constitutive equations
in materials with symmetry,\cite{doc_03433,doc_03434,a1044} physical systems
of high--energy physics\cite{a1019,a1020} and molecular
physics,\cite{a0628,doc_03084,doc_03973} the description of
qubits,\cite{a1477,a1478} $\ldots$

Our previous paper\cite{a1776} considered the complete
permutation--rotation--inversion group of a \methane{} molecule. An integrity
basis for the invariant polynomials was computed. The calculation was
decomposed into two steps and this decomposition was an important feature of
the method. First, we were dealing with the rotation--inversion group
$\mathrm{O(3)}$ and in a second step with the finite permutation group. 
In contrast, in the present paper we are only dealing with finite groups.
The structure of covariants for the rotation--inversion group is interesting on
its own, since it raises specific problems related to the fact that the
modules of covariants are not necessarily free for reductive continuous groups
such as $\mathrm{O(2)}$ or
$\mathrm{O(3)}$.\cite{doc_03189,doc_03190,doc_03181} This is a remarkable
difference with respect to the algebraic structure of invariants. The
non--free modules of $\mathrm{SO(2)}$ have been discussed in \cite{doc_03181}
and a forthcoming article will be devoted to the study of covariant modules of
the $\mathrm{SO(3)}$ group.\cite{doc_03736}

The focus of the present article is on the $\Gamma^\mathrm{final}$--covariants
built from symmetry coordinates for the tetrahedral point group $\Td$, 
although the techniques employed would work for any finite group. 
As a matter of fact, various types of such coordinates have appeared in the
literature for this system that are amenable to our treatment. We can mention
curvilinear internal displacements (bond lengths and interbond
angles),\cite{doc_03239,doc_03729} Cartesian normal
coordinates,\cite{doc_03239,doc_03649,doc_03600,doc_03073,doc_03072,CassamChenai201377}
symmetrized coordinates based on Morse coordinates on Radau vectors for
stretching modes and cosines of valence bond angles for bending modes,
\cite{doc_03720} haversines of bond angles,\cite{Schmidling} cosines of
valence bond angles times functions of bond lengths,\cite{doc_03730}
symmetrized coordinates based on bond lengths, interbond angles and torsion
angles,\cite{doc_03595} or interbond angles and bond lengths times a gaussian
exponential factor.\cite{doc_03538}

% In their approach, they used the Molien series associated to finite molecular point group actions to obtain the dimensions of basis made of homogeneous invariant polynomials of internal coordinates. Since, given a linear representation of a (finite or, more generally, compact) group $G$, all smooth $G$-invariant functions are smooth functions of invariant polynomials (Schwarz, 1975, see e.g. \cite{a0628}), this approach is suitable to express any polynomial, analytic or $C^\infty$ invariant functions. 
The purpose of the present article is to show on the explicit exemple of a
\methane{} molecule that the techniques of invariant theory that have been used to
obtain a polynomial basis set for totally symmetric quantities can be
 extended to quantities transforming according to an
arbitrary irrep.  This is
useful to obtain very efficiently a basis set of $F_2$--symmetry--adapted
polynomials, in the $\Td$-symmetry group,  up to any arbitrary degree, for example.  Such a basis can be used
to fit the EDMS of methane.  The $F_1$--covariants might be relevant to fit
the magnetic dipole moment surface (MDMS) while the $E$--covariants might be
required for components of the quadrupole moment surfaces.  Various
already existing algorithms could theoretically be used for the same purpose
such as those associated to Gr\"obner basis computations.\cite{b164} However,
on the one hand, existing methods of computational invariant
theory\cite{b123,doc_03221,b165} are usually implemented in available computer
codes for invariants only, and on the other hand, they do not seem to be able
to treat high--dimensional problems efficiently for intrinsic complexity
reasons, even in the case of invariants.

The article is organized as follows. In the next section, we recall
fundamental results of invariant theory and illustrate its mathematical
concepts with a case example of $\mathrm{C_i}$ symmetry. Then, we show how the
integrity basis of $\Gamma^\mathrm{final}$--covariant polynomials in the $\Td$
point group can be constructed recursively for \methane{} molecules,
$\Gamma^\mathrm{final}\in\{A_1,A_2,E,F_1,F_2\}$.  The resulting minimal
generating families of symmetry--adapted functions are listed in the
supplemental material.\cite{doc_03735} In conclusion, we emphasize that our approach is
general, as only minor points are specific to the example chosen as an
illustration.

\section{Symmetry--adaptation to a finite group $G$\label{ideas}}

The theoretical framework to describe invariants in polynomial algebras under
finite group actions is well developed, both in mathematics and in chemical
physics. Classical references on the subject in mathematics are the book by
Benson \cite{b098} and the article of Stanley.\cite{a1027} Schmelzer and Murrell
\cite{doc_01779} have had a pioneering influence as far as the construction of
a PES is concerned. The review of Michel and Zhilinski\'{\i}\cite{a0628}
gives an overview of the various possible applications to chemistry and
physics.

We rely in the present section on a fundamental result of commutative algebra
and representation theory stating that any invariant or
$\Gamma^\mathrm{final}$--covariant polynomials has a general decomposition. We
refer to Stanley\cite{a1027} for further details and proofs regarding this
result and other properties of finite group actions on polynomial algebras.

\subsection{Hironaka decomposition\label{section_hironaka}}

Let $\mathcal{P}$ denote the algebra of polynomials in $k$ coordinates,
$Q_1,\ldots,Q_k$, for the field of complex numbers.  This algebra is a direct
sum of vector spaces $\mathcal{P}_n$ of polynomials of degree~$n$:
$\mathcal{P}=\bigoplus\limits_{n\geq 0}\mathcal{P}_{n}$.  We assume that the
finite group $G$ acts linearly on the vector space $<Q_1,\ldots,Q_k>$ spanned
by $Q_1,\ldots,Q_k$. This action extends naturally to $\mathcal{P}$.

Let $\mathcal{P}^{\Gamma^\mathrm{final}} \subset \mathcal{P}$ be the vector
subspace of polynomials transforming as the irrep $\Gamma^\mathrm{final}$ and
let $\left[ \Gamma^\mathrm{final} \right]$ be the dimension of the irrep
$\Gamma^\mathrm{final}$.  This integer equals $1$, $2$ or $3$ for most of the
point groups except for the icosahedral point groups $\mathrm{I}$ and
$\mathrm{I_h}$ where irreps of dimensions $4$ and $5$ occur. A representation
of dimension greater than one is qualified as \textit{degenerate}.  It is
convenient to assume for the forthcoming developments that the representation
$\Gamma^\mathrm{final}$ has a distinguished basis
$\psi^{\Gamma^\mathrm{final},1},...,\psi^{\Gamma^\mathrm{final},\left[\Gamma^\mathrm{final}\right]}$.
A polynomial $\varphi^{\Gamma^\mathrm{final}} \in
\mathcal{P}^{\Gamma^\mathrm{final}}$ is then further decomposed as a sum over
$\left[\Gamma^\mathrm{final}\right]$ polynomials,
\begin{equation}
\varphi^{\Gamma^\mathrm{final}}=\sum\limits_{i=1}^{\left[\Gamma^\mathrm{final}\right]}\varphi^{\Gamma^\mathrm{final},i}, 
\label{comp-decomp}
\end{equation}
each term $\varphi^{\Gamma^\mathrm{final},i}$ behaving as the basis function
$\psi^{\Gamma^\mathrm{final},i}$ under the action of the group $G$, see
e.g. equation (3-187) of Hamermesh.\cite{b064} The symmetry type of the
polynomial $\varphi^{\Gamma^\mathrm{final},i}$ is written $\Gamma^\mathrm{final},i$.  We deduce, 
$\mathcal{P}^{\Gamma^\mathrm{final}}=\bigoplus\limits_{i=1}^{\left[\Gamma^\mathrm{final}\right]}\mathcal{P}^{\Gamma^\mathrm{final},i}$
from the decomposition, eq (\ref{comp-decomp}), for the vector space of
$\Gamma^\mathrm{final}$--covariant polynomials.

An important mathematical result is that there exists exactly $k$ algebraically
independent invariant polynomials $\left\{f_1,\ldots,f_k\right\}$ and
% a finitely generated, free module over the algebra, $\mathbb{R}[f_1,...,f_m]$, spanned by $f_1,\ldots,f_m$, i.e. there exist
 a finite number $p_{\Gamma^\mathrm{final}}$ of linearly independent
 polynomials of symmetry $\Gamma^\mathrm{final},i$: 
 $\left\{g^{\Gamma^\mathrm{final},i}_1 , \ldots ,
 g^{\Gamma^\mathrm{final},i}_{p_{\Gamma^\mathrm{final}}}\right\}$, such that
\begin{equation}
\mathcal{P}^{\Gamma^\mathrm{final},i}=
\bigoplus_{j=1}^{p_{\Gamma^\mathrm{final}}}
\mathbb{C}[f_1,\ldots,f_k]g^{\Gamma^\mathrm{final},i}_j,\,
i \in \left\{1,2,\ldots,\left[ \Gamma^\mathrm{final} \right]\right\},
\label{Hironaka}
\end{equation} 
where $\mathbb{C}[f_1,...,f_k]$ is the algebra spanned by the polynomials
$\left\{f_1,\ldots,f_k\right\}$. The number $p_{\Gamma^\mathrm{final}}$
depends on the irrep $\Gamma^\mathrm{final}$ but is independent on the index
$i$.  We refer to the whole set
$\{f_1,...,f_k;g^{\Gamma^\mathrm{final},i}_1,...,g^{\Gamma^\mathrm{final},i}_{p_{\Gamma^\mathrm{final}}}\}$
as an integrity basis for the module $\mathcal{P}^{\Gamma^\mathrm{final},i}$.
The $f_i$ are called the \textit{denominator} or \textit{primary} polynomials,
while the $g^{\Gamma^\mathrm{final}}_j$ are called the \textit{numerator} or
\textit{secondary} polynomials. The same set of primary invariants is used for
all the irreps. Such a decomposition as in eq~\ref{Hironaka} is sometimes
referred to as an Hironaka decomposition and defines a so--called
Cohen--Macaulay module.  In the particular case where $\Gamma^\mathrm{final}$
is the trivial representation (so that $\Gamma^\mathrm{final}$--covariants are
simply invariants), this result shows that
$\mathcal{P}^{\Gamma^\mathrm{final}}$, the algebra of invariant polynomials,
is a Cohen--Macaulay algebra.

The elements of the integrity basis can always be chosen homogeneous, and
from now on, we assume that this homogeneity property always holds. Even with
this assumption, the number of basis polynomials is not determined by the
above construction.  However, for a given choice of primary invariants, the
number of $\Gamma^\mathrm{final}$--covariant basis polynomials and their
degrees are fixed and determined by the so--called Molien series.\cite{a1743}
% Notice that the $\Gamma^\mathrm{final}$-covariant numerator
% polynomials depend also on the choice of a particular basis
% $\psi_1,...,\psi_{n_{\Gamma^\mathrm{final}}}$ for the irrep
% $\Gamma^\mathrm{final}$.
The problem of constructing polynomials of symmetry type
$\Gamma^\mathrm{final},i$ from symmetrized coordinates spanning the
representation $\Gamma^\mathrm{initial}$ leads to consider the Molien series,
$M^{G}\left(\Gamma^\mathrm{final};\Gamma^\mathrm{initial};t\right)$, defined
by:
\begin{equation}
M^{G}\left(\Gamma^\mathrm{final};\Gamma^\mathrm{initial};t\right)
=
\sum\limits_{n\geq0}\dim\mathcal{P}^{\Gamma^\mathrm{final},i}_n\ t^n,
\label{def_Molien_series}
\end{equation} 
where $\mathcal{P}^{\Gamma^\mathrm{final},i}_n =
\mathcal{P}^{\Gamma^\mathrm{final},i}\cap\mathcal{P}_n$ is the vector space of
polynomials of symmetry type $\Gamma^\mathrm{final},i$ and of degree $n$.  In
other words, the coefficient $\dim\mathcal{P}^{\Gamma^\mathrm{final},i}_n$ of
the Molien series gives the number of linearly independent polynomials of
symmetry type $\Gamma^\mathrm{final},i$ and of degree $n$.

Suppose that
$\{f_1,...,f_k;g^{\Gamma^\mathrm{final},i}_1,...,g^{\Gamma^\mathrm{final},i}_{p_{\Gamma^\mathrm{final}}}\}$
is an integrity basis for $\mathcal{P}^{\Gamma^\mathrm{final},i}$. Then it can
be shown that the corresponding Molien series can be cast in the following
form:
\begin{equation}
M^{G}\left(\Gamma^\mathrm{final};\Gamma^\mathrm{initial};t\right)
=
\frac{t^{\deg(g^{\Gamma^\mathrm{final},i}_1)}+\cdots+t^{\deg(g^{\Gamma^\mathrm{final},i}_{p_{\Gamma^\mathrm{final}}})}}
{(1-t^{\deg(f_1)})\cdots(1-t^{\deg(f_k)})},
\label{molien1}
\end{equation}
where $\deg\left(p\right)$ is the degree of the polynomial $p$ (the degrees
are not necessarily all distinct in this expression).  The expression of the
Molien function
$M^{G}\left(\Gamma^\mathrm{final};\Gamma^\mathrm{initial};t\right)$ is
independent of the choice of the index $i$.  The right--hand side of
eq~\ref{molien1} justifies the alternative denomination of the $f_i$ primary
polynomials as \textit{denominator} polynomials and of the
$g^{\Gamma^\mathrm{final},i}_j$ secondary polynomials as \textit{numerator}
polynomials. Once the degrees of the denominator invariants are given and the
Molien function calculated, the number of numerator polynomials of each degree
is given by the corresponding coefficient in the polynomial
$M^{G}\left(\Gamma^\mathrm{final};\Gamma^\mathrm{initial};t\right)\times
(1-t^{\deg(f_1)})\cdots(1-t^{\deg(f_k)})$.  The problem of generating the
module $\mathcal{P}^{\Gamma^\mathrm{final},i}$ comes down to the computation
of a complete set of such numerator polynomials given a set of denominator
invariants.

%\subsection{The mixing principle}
\subsection{Recursive construction \label{recursion}}
\subsubsection{Generating function}
We considered in the previous section the action of a finite group $G$ on a
polynomial algebra $\mathcal{P}$ over a vector space $<Q_1,\ldots,Q_k>$.  In
our applications of invariant theory, the representation
$\Gamma^\mathrm{initial}$ spanned by the symmetrized coordinates typically
splits into a direct sum of $\mu$ irreps
$\Gamma^\mathrm{initial}_i$, $1 \leq i \leq \mu$,
$$
\Gamma^\mathrm{initial}
=
\bigoplus_{i=1}^\mu \Gamma^\mathrm{initial}_i.
$$
The definition of the Molien series in eq~\ref{def_Molien_series} of
Section~\ref{section_hironaka} involved only one variable $t$.  In order to
follow the contributions of the different irreps $\Gamma^\mathrm{initial}_i$,
we introduce now one $t_i$ variable for each $\Gamma^\mathrm{initial}_i$ and
write $M^{G}\left(\Gamma;\Gamma^\mathrm{initial}_1 \oplus
\Gamma^\mathrm{initial}_2 \oplus \cdots \oplus
\Gamma^\mathrm{initial}_k;t_1,t_2,\ldots,t_k\right)$ for the Molien series
associated to $\Gamma$--covariants polynomials in the variables
contained in the reducible irrep $\Gamma^\mathrm{initial}_1 \oplus
\Gamma^\mathrm{initial}_2 \oplus \cdots \oplus \Gamma^\mathrm{initial}_k$
under group $G$.

Let us note $c_{\Gamma_\alpha,\Gamma_\beta}^{\Gamma}$ for the multiplicity of the
irrep $\Gamma$ in the direct (or Kronecker) product $\Gamma_\alpha
\times \Gamma_\beta$ of the irreps $\Gamma_\alpha$ and
$\Gamma_\beta$.  In case of the $\Td$ point group,
$c_{\Gamma_\alpha,\Gamma_\beta}^{\Gamma}=0$ or $1$, see Wilson \textit{et
  al.}.\cite{b101} Decomposing the initial reducible representation
$\Gamma^\mathrm{initial}$ as
$$
\Gamma^\mathrm{initial}
=
\left(
\Gamma^\mathrm{initial}_1\oplus\cdots\oplus\Gamma^\mathrm{initial}_{\mu-1}
\right)
\oplus
\Gamma^\mathrm{initial}_\mu,
$$
(note the parentheses), the generating function
$M^{G}\left(\Gamma;\Gamma^\mathrm{initial};t_1,t_2,\ldots,t_\mu\right)$ can be
built by coupling the generating functions
$$
M^G\left(\Gamma_\alpha;\Gamma^\mathrm{initial}_1 \oplus \Gamma^\mathrm{initial}_2
\oplus \cdots \oplus \Gamma^\mathrm{initial}_{\mu-1};
t_1,t_2,\ldots,t_{\mu-1}\right),
$$
with the generating functions
$$
M^{G}\left(\Gamma_\beta;\Gamma^\mathrm{initial}_\mu;t_\mu\right),
$$
where $\Gamma_\alpha$ and $\Gamma_\beta$ are irreps, (see
Equation~(46) of Michel and Zhilinski\'{\i}\cite{a0628} and
Appendix~\ref{GF_IB_TD},) according to the following equation:
\begin{eqnarray}
\lefteqn{M^{G}\left(\Gamma;\Gamma^\mathrm{initial}_1\oplus\Gamma^\mathrm{initial}_2\oplus\cdots\oplus
\Gamma^\mathrm{initial}_{\mu-1}\oplus\Gamma^\mathrm{initial}_\mu;t_1,t_2,\ldots,t_{\mu-1},t_\mu\right)=}\nonumber\\
&\sum\limits_{\Gamma_\alpha,\Gamma_\beta}c_{\Gamma_\alpha,\Gamma_\beta}^{\Gamma}
M^{G}\left(\Gamma_\alpha;
\Gamma^\mathrm{initial}_1 \oplus \Gamma^\mathrm{initial}_2 \oplus
\cdots \oplus \Gamma^\mathrm{initial}_{\mu-1};
t_1,t_2,\ldots,t_{\mu-1}\right)
\times
M^{G}\left(\Gamma_\beta;\Gamma^\mathrm{initial}_\mu;t_\mu\right).
\label{molienmixed}
\end{eqnarray}
In eq~\ref{molienmixed}, the double sum on $\Gamma_\alpha$ and $\Gamma_\beta$
runs over all the irreps of the group $G$.  The Molien function
$M^{G}\left(\Gamma_\alpha;\Gamma^\mathrm{initial}_1\oplus\Gamma^\mathrm{initial}_2\oplus\cdots\oplus\Gamma^\mathrm{initial}_{\mu-1};t_1,t_2,\ldots,t_{\mu-1}\right)$
in the right--hand side of eq~\ref{molienmixed} can itself be computed through
an equation similar to eq~\ref{molienmixed} if the representation
$\Gamma^\mathrm{initial}_1\oplus\Gamma^\mathrm{initial}_2\oplus\cdots\oplus\Gamma^\mathrm{initial}_{\mu-1}$
is seen as a direct sum of
$\Gamma^\mathrm{initial}_1\oplus\Gamma^\mathrm{initial}_2
\oplus\cdots\oplus\Gamma^\mathrm{initial}_{\mu-2}$ and
$\Gamma^\mathrm{initial}_{\mu-1}$.  These iterations are continued until no
more decomposition of the representations is possible.  The left--hand side of
eq~\ref{molienmixed} is then ultimately written as a sum of products of
\textit{elementary generating functions}
$M^{G}\left(\Gamma_\alpha;\Gamma^\mathrm{initial}_i;t_i\right)$ where both
$\Gamma_\alpha$ and $\Gamma^\mathrm{initial}_i$ are irreps. Such elementary
generating functions have already appeared in the litterature for a variety of
 point groups\cite{a1392}, (see also Appendix~\ref{GF_IB_TD}).  These
elementary generating functions are the building blocks required to  compute recursively
according to  eq~\ref{molienmixed}, the Molien generating function of the problem under study.

\subsubsection{Integrity basis\label{recursion_ib}}

To each generating function of the form, eq~\ref{molien1}, correspond  integrity bases whose
 number and degree of the denominator and numerator polynomials are suggested by such an expression.  
Let
$$\bigcup_{x\in \left\{1,2,\cdots,\left[\Gamma_\alpha\right]\right\}}
\{f_1,...,f_k;g^{\Gamma_\alpha,x}_1,...,g^{\Gamma_\alpha,x}_{p_{\Gamma_\alpha}}\},
$$
be an integrity basis corresponding to the generating function
\begin{equation}
M^G\left(\Gamma_\alpha;\Gamma^\mathrm{initial}_1 \oplus \cdots \oplus
\Gamma^\mathrm{initial}_{i-1};t_1,\ldots,t_{i-1}\right),
\label{genfun_a}
\end{equation}
 and let
$$
\bigcup_{y\in \left\{1,2,\cdots,\left[\Gamma_\beta\right]\right\}}
\{h_1,...,h_l;j^{\Gamma_\beta,y}_1,...,j^{\Gamma_\beta,y}_{p_{\Gamma_\beta}}\},
$$
be an integrity basis corresponding to the generating function
\begin{equation}
M^G\left(\Gamma_\beta;\Gamma^\mathrm{initial}_i;t_i\right).
\label{genfun_b}
\end{equation}
The $f_i$ and $g^{\Gamma_\alpha,x}_j$ are polynomials in the variables of the
representation $\Gamma^\mathrm{initial}_1 \oplus \cdots \oplus
\Gamma^\mathrm{initial}_{i-1}$, while the $h_i$ and $j^{\Gamma_\beta,y}_j$ are
polynomials in the variables of the representation $\Gamma^\mathrm{initial}_i$.

The set $\{f_1,...,f_k, h_1,...,h_l\}$ is the set of denominator or primary
invariants for the generating function
\begin{equation}
M^G\left(\Gamma;\Gamma^\mathrm{initial}_1\oplus\cdots\oplus\Gamma^\mathrm{initial}_{i-1}
\oplus \Gamma^\mathrm{initial}_i;t_1,\ldots,t_{i-1},t_i\right).
\label{our_method}
\end{equation}
  The numerator or secondary
polynomials of the generating function of eq~\ref{our_method} are generated by
coupling the numerator polynomials $g^{\Gamma_\alpha,1}_a$, \ldots,
$g^{\Gamma_\alpha,\left[\Gamma_\alpha\right]}_a$ with the numerator
polynomials $j^{\Gamma_\beta,1}_b$, \ldots,
$j^{\Gamma_\beta,\left[\Gamma\beta\right]}_b$ via the Clebsch--Gordan
coefficients of the group $G$ for all $\left(\Gamma_\alpha,\Gamma_\beta\right)$
pairs such that $\Gamma \in \Gamma_\alpha \times
\Gamma_\beta$, see Section 5.6 of Hamermesh.~\cite{b064} We write these
functions $m^{\Gamma,\kappa}_{\Gamma_\alpha,\Gamma_\beta,a,b,i}$, where $1\leq i\leq
c_{\Gamma_\alpha,\Gamma_\beta}^{\Gamma}$, $1\leq a\leq p_{\Gamma_\alpha}$, $1\leq b\leq
p_{\Gamma_\beta}$, and $\kappa \in \left\{1,2,\cdots,\left[
  \Gamma \right]\right\}$. The resulting integrity basis corresponding to 
eq~\ref{our_method} can be expressed as 
\begin{tiny}$$\{f_1,...,f_k,h_1,...,h_l;\bigcup\limits_{\Gamma_\alpha,\Gamma_\beta}\{m^{\Gamma,\kappa}_{\Gamma_\alpha,\Gamma_\beta,a,b,i},
1\leq a\leq p_{\Gamma_\alpha}, 
1\leq b\leq p_{\Gamma_\beta},
1\leq i\leq c_{\Gamma_\alpha,\Gamma_\beta}^{\Gamma},
\kappa \in \left\{1,2,\cdots,\left[ \Gamma \right]\right\}\},$$\end{tiny}
(if  $\Gamma \notin \Gamma_\alpha \times\Gamma_\beta$, $c_{\Gamma_\alpha,\Gamma_\beta}^{\Gamma}=0$, 
and the set of $m^{\Gamma,\kappa}_{\Gamma_\alpha,\Gamma_\beta,a,b,i}$'s is empty).

% Eq~\ref{molienmixed} contains a double sum over $\Gamma_\alpha$ and $\Gamma_\beta$.
% Accordingly, we must consider for the construction of the integrity basis corresponding to
% eq~\ref{our_method} all $\left(\Gamma_\alpha,\Gamma_\beta\right)$
% pairs such that their direct product contains $\Gamma$. 
% Such an integrity basis  is given by the set of denominator polynomials
% $\{f_1,...,f_k,h_1,...,h_l\}$ and the set of numerator
% polynomials
% $\bigcup\limits_{\Gamma_\alpha,\Gamma_\beta}\{m^{\Gamma,\kappa}_{\Gamma_\alpha,\Gamma_\beta,a,b,i},
% a\leq p_{\Gamma_\alpha}, 
% b\leq p_{\Gamma_\beta},
% i\leq c_{\Gamma_\alpha,\Gamma_\beta}^{\Gamma},
% \kappa \in \left\{1,2,\cdots,\left[ \Gamma_\mathrm{final} \right]\right\}$.

So, the integrity basis is built in a straightforward manner
from integrity bases associated to generating functions eqs (\ref{genfun_a}) and
(\ref{genfun_b}), where both initial representations are of smaller dimensions.
% We have thus an algorithm to built up the required integrity basis from
% smaller representations.
%  In particular, the integrity bases for 
Iterating this process constitutes an effective algorithm which only needs 
the elementary generating functions of group G for its initialization. 
The latter functions have already been
tabulated\cite{a1392} for most groups of interest. 
The algorithm terminates when all $ \Gamma^\mathrm{initial}_i$'s have been incorporated.
% and can be used as building blocks to recursively
% determine the integrity basis associated with the left--hand side of
% eq~\ref{molienmixed} for a given $\Gamma=\Gamma^\mathrm{final}$ irrep.  At the
% end of the computation, we get the primary and secondary polynomials of the
% integrity basis associated with the generating function of
% eq~\ref{our_method}. 
%As already mentionned, an integrity basis for $\mathcal{P}^{[2]^{\Gamma}}$ is easily generated from one of $\mathcal{P}^{[2]^{\Gamma,1}}$ see Ref.~\cite[Sect. 3-18]{doc_03731}.
%The process is iterated by substituting
%$\mathcal{P}^{[2]^{\Gamma_\alpha}}$ to $\mathcal{P}_{a_1}^{\Gamma_\alpha}$ and
%$\mathcal{P}_{a_3}^{\Gamma_\beta}$ to $\mathcal{P}_{a_2}^{\Gamma_\beta}$, and so on
%recursively. 
% However, for notational simplicity, we have described only the
% first step of the recursion.

\subsection{Illustration on a case example
\label{tutorial}}
The present section gives a straightforward application of the recursive
construction in the simplest non trivial case of the two-element group, which can be taken 
as the $\mathrm{C_i}$ group used in chemistry for molecular structures with a center of inversion.
\subsubsection{Group $\mathrm{C_i}$}
  The group
$\mathrm{C_i}$ has two elements: the identity operation $E$ leaves unchanged
the coordinates of the particles, $x \mapsto x$, while the inversion operation
$I$ changes the sign of the coordinates, $x \mapsto -x$.  The character table
of the $\mathrm{C_i}$ group is given in Table~\ref{Ci_character_table} and
shows that two one--dimensional irreps $A_1$ and $A_2$ occur in this group.

\begin{table}[h!]
\begin{center}
\caption{Character table of the $\mathrm{C_i}$ point group.
\label{Ci_character_table}}
\begin{tabular}{ccc}
\hline \hline
& $E$ & $I$ \\
\hline
$A_1$ & $1$ & $1$ \\
$A_2$ & $1$ & $-1$ \\
\hline \hline
\end{tabular}
\end{center}
\end{table}

\subsubsection{Elementary generating functions}

Applications of group theory often search to construct objects that transforms
as a final irrep $\Gamma^\mathrm{final}$ of a group $G$ from elementary
objects that spans an initial, possibly reducible, representation
$\Gamma^\mathrm{initial}$. If these objects are polynomials, we can sort them
by their degree and count the number $c_k$ of linearly independent polynomials
of degree $k$ that can be built up. The information on the $c_k$'s is encoded
into the so--called Molien series or generating function:
\begin{equation}
M^G\left(\Gamma^\mathrm{final};\Gamma^\mathrm{initial};t\right)
=
c_0+c_1t+c_2t^2+c_3t^3+\cdots.
\label{Molien_series}
\end{equation}

Elementary generating functions are particular generating functions when both
the initial representation $\Gamma^\mathrm{initial}$ and the final
representation $\Gamma^\mathrm{final}$ are irreps of the group.  The group
$\mathrm{C_i}$ has two irreps and thus four elementary generating functions
have to be considered: $M^{\mathrm{C_i}}\left(A_1;A_1;t\right)$,
$M^{\mathrm{C_i}}\left(A_2;A_1;t\right)$,
$M^{\mathrm{C_i}}\left(A_1;A_2;t\right)$, and
$M^{\mathrm{C_i}}\left(A_2;A_2;t\right)$.

\subsubsection{$M^{\mathrm{C_i}}\left(\Gamma^\mathrm{final};A_1;t\right)$}
The absolute value $\left|x\right|$ is a good example of an $A_1$--symmetric (invariant)
object as it does not change sign under neither the identity $E$ nor the
inversion $I$
operations. From $\left|x\right|$ can be constructed one invariant of degree
$0$ ($\left|x\right|^0=1$), one invariant of degree $1$ ($\left|x\right|^1$), 
one invariant of degree two ($\left|x\right|^2$), and more generally, one
invariant of degree $k$ ($\left|x\right|^k$). However, no object of symmetry
$A_2$ can be constructed from $\left|x\right|$. As a consequence, we can write
as in eq~\ref{Molien_series} the expressions of the Molien series
 $M^{\mathrm{C_i}}\left(\Gamma^\mathrm{final};A_1;t\right)$:
\begin{equation}
\begin{array}{rclcl}
M^{\mathrm{C_i}}\left(A_1;A_1;t\right)
&=&
1+t+t^2+t^3+t^4+\cdots
&=&
\frac{1}{1-t}, \\
M^{\mathrm{C_i}}\left(A_2;A_1;t\right)
&=&
0.
\end{array}
\label{appendix_1_005}
\end{equation}

\subsubsection{$M^{\mathrm{C_i}}\left(\Gamma^\mathrm{final};A_2;t\right)$}
The monomial $x$ is an example of an $A_2$--symmetric object because it changes
sign under the inversion $I$ operation. The even powers of $x$ will be of
$A_1$--symmetry:
\begin{equation}
x^{2n} \mapsto \left(-x\right)^{2n}=x^{2n},
\label{even}
\end{equation}
while the odd powers of $x$ will be of
$A_2$--symmetry:
\begin{equation}
x^{2n+1} \mapsto \left(-x\right)^{2n+1}=-x^{2n+1}.
\label{odd}
\end{equation}
We see that from an $A_2$--symmetric object can be constructed
one object of symmetry $A_1$ of any even degree  and one object of symmetry
$A_2$ of any odd degree. These results are encoded in the two following Molien series:
\begin{equation}
\begin{array}{rclcl}
M^{\mathrm{C_i}}\left(A_1;A_2;t\right)
&=&
1+t^2+t^4+t^6+\cdots
&=&
\frac{1}{1-t^2}, \\
M^{\mathrm{C_i}}\left(A_2;A_2;t\right)
&=&
t+t^3+t^5+t^7+\cdots
&=&
\frac{t}{1-t^2}.
\end{array}
\label{appendix_1_008}
\end{equation}

\subsubsection{Integrity bases for the elementary generating functions}
An integrity basis consists in two sets of polynomials, the denominator and
the numerator polynomials.  A generating function written as in the
right--hand side of eq~\ref{molien1} suggests both the number and the degree
of the denominator and numerator polynomials, and is a very valuable source of
information when an integrity basis is built up. When forming polynomials that
transform as the $\Gamma^\mathrm{final}$ irrep from polynomials that belongs
to the integrity basis corresponding to 
$M^G\left(\Gamma^\mathrm{final};\Gamma^\mathrm{initial};t\right)$,
eq~\ref{Hironaka} indicates that the denominator polynomials can be multiplied
between them with no restriction at all while the numerator polynomials only
appear linearly. The explicit expressions of the integrity bases for the four
elementary generating functions of the group $\mathrm{C_i}$ are given in
Table~\ref{integrity_bases_egf}. For example, the last line of
Table~\ref{integrity_bases_egf} suggests that we can recover all the
polynomials of symmetry $A_2$ built up from $x$ by multiplying the numerator
polynomial $x$ with any power of the denominator polynomial $x^2$. The final
result is a polynomial of the form $x^{2n+1}$ which has the desired symmetry,
see eq~\ref{odd}.

\begin{table}[h!]
\begin{center}
\caption{Integrity bases for the four elementary generating functions of the
  group $\mathrm{C_i}$. 
\label{integrity_bases_egf}}
\begin{tabular}{lll}
\hline \hline
Generating function & Denominator polynomials & Numerator polynomials \\
\hline
$M^{\mathrm{C_i}}\left(A_1;A_1;t\right)=\frac{1}{1-t}$ &
$\left\{\left|x\right|\right\}$ & $\left\{1\right\}$ \\
$M^{\mathrm{C_i}}\left(A_2;A_1;t\right)=0$ & & \\
$M^{\mathrm{C_i}}\left(A_1;A_2;t\right)=\frac{1}{1-t^2}$ &
$\left\{x^2\right\}$ & $\left\{1\right\}$ \\
$M^{\mathrm{C_i}}\left(A_2;A_2;t\right)=\frac{t}{1-t^2}$ &
$\left\{x^2\right\}$ & $\left\{x\right\}$ \\
\hline \hline
\end{tabular}
\end{center}
\end{table}

\subsubsection{Case example}

Let us consider three particles moving on an infinite straight line under the
symmetry group $\mathrm{C_i}$. The position of the three particles are given
by $x_i$, $1 \leq i \leq 3$. The action of the inversion $I$ changes the
coordinates of the three particles: $x_i \mapsto -x_i$.  The three $x_i$
variables can be seen as polynomials of degree one. They are manifestly of
symmetry $A_2$, hence the initial reducible representation is
$\Gamma^\mathrm{initial}=A_2 \oplus A_2 \oplus A_2$.  Polynomials of higher
degree can be built up from the $x_1$, $x_2$ and $x_3$ polynomials and the
example is simple enough that the symmetry of the higher degree polynomials is
immediately deduced.

The case example is to construct all the polynomials in $x_1$, $x_2$, and $x_3$
of symmetry $A_1$ or $A_2$ up to a given degree. This is the kind of problem
that appear when the potential energy surface or the electric dipole moment
surface are expanded in symmetry--adapted
polynomials. Table~\ref{higher_degree_polynomials} gives a list of the
linearly independent polynomials of low degree in $x_1$, $x_2$, and $x_3$ that
can be found by manual inspection.

% mtaylor((1+x1*x2+x1*x3+x2*x3)/((1-x1^2)*(1-x2^2)*(1-x3^2)),[x1=0,x2=0,x3=0],5);
% mtaylor((x1+x2+x3+x1*x2*x3)/((1-x1^2)*(1-x2^2)*(1-x3^2)),[x1=0,x2=0,x3=0],6);
\begin{table}[h!]
\begin{center}
\caption{Linearly independent polynomials of degree $k$, $0 \leq k\leq 5$, in
  variables $x_1$, $x_2$, and $x_3$ transforming according to the
  irrep $\Gamma^\mathrm{final}$. The number of such
  polynomials is noted $\dim\mathcal{P}^{\Gamma^\mathrm{final}}_k$. 
\label{higher_degree_polynomials}}
\begin{tabular}{ccp{10cm}c}
\hline \hline
$\Gamma^\mathrm{final}$ & $k$ & Polynomials & $\dim\mathcal{P}^{\Gamma^\mathrm{final}}_k$ \\
\hline
$A_1$ & $0$ & $1$ & $1$ \\
$A_1$ & $2$ & $x_1^2$, $x_1x_2$, $x_1x_3$, $x_2^2$, $x_2x_3$, $x_3^2$ & $6$ \\
$A_1$ & $4$ & $x_1^4$, $x_1^3x_2$, $x_1^3x_3$, $x_1^2x_2^2$, $x_1^2x_2x_3$,
$x_1^2x_3^2$, $x_1x_2^3$, $x_1x_2^2x_3$, $x_1x_2x_3^2$, $x_1x_3^3$, $x_2^4$,
$x_2^3x_3$, $x_2^2x_3^2$, $x_2x_3^3$, $x_3^4$ & $15$ \\
\hline
$A_2$ & $1$ & $x_1$, $x_2$, $x_3$ & $3$ \\
$A_2$ & $3$ & $x_1^3$, $x_1^2x_2$, $x_1^2x_3$, $x_1x_2^2$, $x_1x_2x_3$,
$x_1x_3^2$, $x_2^3$, $x_2^2x_3$, $x_2x_3^2$, $x_3^3$ & $10$ \\
$A_2$ & $5$ & $x_1^5$, $x_1^4x_2$, $x_1^4x_3$, $x_1^3x_2^2$, $x_1^3x_2x_3$, $x_1^3x_3^2$, 
$x_1^2x_2^3$, $x_1^2x_2^2x_3$, $x_1^2x_2x_3^2$, $x_1^2x_3^3$,
$x_1x_2^4$, $x_1x_2^3x_3$, $x_1x_2^2x_3^2$, $x_1x_2x_3^3$, $x_1x_3^4$, 
$x_2^5$, $x_2^4x_3$, $x_2^3x_3^2$, $x_2^2x_3^3$,  $x_2x_3^4$, $x_3^5$
& $21$ \\
\hline \hline
\end{tabular}
\end{center}
\end{table}
% taylor((1+3*t^2)/((1-t^2)^3),t=0,5);
% taylor((3*t+t^3)/((1-t^2)^3),t=0,6);

From the last column of Table~\ref{higher_degree_polynomials} and remembering that the
coefficient $c_k$ in eq~\ref{Molien_series} is the number of linearly independent polynomials of
degree $k$ for a given final symmetry, the generating functions are found to be:
\begin{equation}
M^{\mathrm{C_i}}\left(A_1;\Gamma^\mathrm{initial};t\right)
=1+6t^2+15t^4+\cdots,
\label{appendix_1_001}
\end{equation}
for the $\Gamma^\mathrm{final}=A_1$ representation, and
\begin{equation}
M^{\mathrm{C_i}}\left(A_1;\Gamma^\mathrm{initial};t\right)
=3t+10t^3+21t^5+\cdots,
\label{appendix_1_002}
\end{equation}
or the $\Gamma^\mathrm{final}=A_2$ representation.

These generating functions can be directly computed using the Molien's formula
and Burnside's generalization to final irrep different from the totally
symmetry one.\cite{a1743,b167} For a finite point group $G$, the Molien
function reads:
\begin{equation}
M^G\left(\Gamma^\mathrm{final};\Gamma^\mathrm{initial};t\right) 
=
\frac{1}{\left|G\right|}
\sum_{g\in G} \frac{\bar{\chi}\left(\Gamma^\mathrm{final};g\right)}
{\det \left( 1_{n\times n} - t M\left(\Gamma^\mathrm{initial};g\right)
  \right)},
\label{Molien_formula}
\end{equation}
where 
%the sum is over all the elements $g$ of the group, 
$\left|G\right|$ is the order of $G$, 
$\bar{\chi}\left(\Gamma^\mathrm{final};g\right)$ is the complex
conjugate of the character for element $g\in G$ and irrep
$\Gamma^\mathrm{final}$, $1_{n\times n}$ is the $n \times n$ identity matrix acting on $\Gamma^\mathrm{initial}$ of dimension $n$,
$M\left(\Gamma^\mathrm{initial};g\right)$ is the $n\times n$ matrix
representation of $g$ on $\Gamma^\mathrm{initial}$, and $\det$ is the determinant of a matrix.

In our example, the representation matrices of the
$\Gamma^\mathrm{initial}=A_2 \oplus A_2 \oplus A_2$ are the two $3\times 3$
diagonal matrices: $M\left(\Gamma^\mathrm{initial};E\right)=\mathrm{diag}
\left(1,1,1\right)$ and $M\left(\Gamma^\mathrm{initial};I\right)=\mathrm{diag}
\left(-1,-1,-1\right)$.  Using Table~\ref{Ci_character_table}, the
representation matrices and Molien's formula \ref{Molien_formula}, we find the
two generating functions:
\begin{eqnarray}
M^{\mathrm{C_i}}\left(A_1;\Gamma^\mathrm{initial};t\right)
&=&
\frac{1+3t^2}{\left(1-t^2\right)^3},
\label{appendix_1_003} \\
M^{\mathrm{C_i}}\left(A_2;\Gamma^\mathrm{initial};t\right)
&=&
\frac{3t+t^3}{\left(1-t^2\right)^3}.
\label{appendix_1_004}
\end{eqnarray}
It can be checked that the Taylor series of eqs~\ref{appendix_1_003} and
\ref{appendix_1_004} around $t=0$ correspond to the expansion whose beginning is given
in eqs~\ref{appendix_1_001} and \ref{appendix_1_002}.  The generating
function eq~\ref{appendix_1_003} suggests that the integrity basis for the
invariants built from $x_1$, $x_2$, and $x_3$ consists of three denominator
polynomials of degree two and four numerator polynomials, of which one is of
degree zero and three are of degree two.  Eq~\ref{appendix_1_004} suggests
that the integrity basis for the polynomials of symmetry $A_2$ consists of
three denominator polynomials of degree two and four numerator polynomials, of
which three are of degree one and one is of degree three. 

\subsubsection{Recursive construction of the generating functions}
Eqs~\ref{appendix_1_003} and \ref{appendix_1_004} were obtained from Molien's formula \ref{Molien_formula}.
However, they can be derived more efficiently from the recursive construction of section~\ref{recursion}. 

Let us use eq~\ref{molienmixed}  to compute recursively
the generating functions for our case example from the elementary generating
functions of $\mathrm{C_i}$.  
%First, eq~\ref{molienmixed} is applied to our
% problem.  The $c_{\Gamma_\alpha,\Gamma_\beta}^{\Gamma}$ is the multiplicity of
% the irrep in the direct product of the irreps $\Gamma_\alpha$ and
% $\Gamma_\beta$. 
Noting the direct products
$A_1\times A_1=A_2\times A_2=A_1$ and $A_1\times A_2=A_2\times A_1=A_2$, only
four $c_{\Gamma_\alpha,\Gamma_\beta}^{\Gamma}$ coefficients do not vanish:
\begin{equation}
c_{A_1,A_1}^{A_1}=c_{A_2,A_2}^{A_1}=c_{A_1,A_2}^{A_2}=c_{A_2,A_1}^{A_2}=1.
\label{direct_products_ci}
\end{equation}
The generating function for the invariant polynomials in $x_1$,
$x_2$, and $x_3$ is, according to eqs~\ref{molienmixed} and
\ref{direct_products_ci}:
\begin{eqnarray}
\lefteqn{M^{\mathrm{C_i}}\left(A_1;A_2 \oplus A_2 \oplus A_2 ; t_1,t_2,t_3 \right)}
\nonumber \\
&=&
M^{\mathrm{C_i}}\left(A_1;A_2 \oplus A_2 ; t_1,t_2 \right)
M^{\mathrm{C_i}}\left(A_1;A_2 ; t_3 \right)
\nonumber \\
&&+
M^{\mathrm{C_i}}\left(A_2;A_2 \oplus A_2 ; t_1,t_2 \right)
M^{\mathrm{C_i}}\left(A_2;A_2 ; t_3 \right).
\nonumber
\end{eqnarray}
Each of the $M^{\mathrm{C_i}}\left(\Gamma_\alpha;A_2 \oplus A_2 ; t_1,t_2 \right)$ term
can again be decomposed using eq~\ref{molienmixed}, and we finally find a
relation where only elementary generating functions appear in the right--hand side:
\begin{eqnarray}
\lefteqn{M^{\mathrm{C_i}}\left(A_1;A_2 \oplus A_2 \oplus A_2 ; t_1,t_2,t_3 \right)}
\nonumber \\
&=&
M^{\mathrm{C_i}}\left(A_1;A_2; t_1 \right)
M^{\mathrm{C_i}}\left(A_1;A_2; t_2 \right)
M^{\mathrm{C_i}}\left(A_1;A_2; t_3 \right)
\nonumber \\
&&+
M^{\mathrm{C_i}}\left(A_2;A_2; t_1 \right)
M^{\mathrm{C_i}}\left(A_2;A_2; t_2 \right)
M^{\mathrm{C_i}}\left(A_1;A_2; t_3 \right)
\nonumber \\
&&+
M^{\mathrm{C_i}}\left(A_1;A_2; t_1 \right)
M^{\mathrm{C_i}}\left(A_2;A_2; t_2 \right)
M^{\mathrm{C_i}}\left(A_2;A_2; t_3 \right)
\nonumber \\
&&+
M^{\mathrm{C_i}}\left(A_2;A_2; t_1 \right)
M^{\mathrm{C_i}}\left(A_1;A_2; t_2 \right)
M^{\mathrm{C_i}}\left(A_2;A_2; t_3 \right).
\label{appendix_1_009}
\end{eqnarray}

The expressions of the elementary generating functions are given in
eqs~\ref{appendix_1_005} and~\ref{appendix_1_008}, and the expression for the
invariants reads as:
\begin{equation}
M^{\mathrm{C_i}}\left(A_1;A_2 \oplus A_2 \oplus A_2 ; t_1,t_2,t_3 \right)
=\frac{1+t_1t_2+t_1t_3+t_2t_3}{\left(1-t_1^2\right)\left(1-t_2^2\right)\left(1-t_3^2\right)}.
\label{appendix_1_010}
\end{equation}
If the three $A_2$ in the initial reducible representation are not
distinguished, we can write $t_1=t_2=t_3=t$ in eq~\ref{appendix_1_010} to recover eq~\ref{appendix_1_003}. 
The same method for $\Gamma^\mathrm{final}=A_2$
gives
\begin{equation}
M^{\mathrm{C_i}}\left(A_2;A_2 \oplus A_2 \oplus A_2 ; t_1,t_2,t_3 \right)
=\frac{t_1+t_2+t_3+t_1t_2t_3}{\left(1-t_1^2\right)\left(1-t_2^2\right)\left(1-t_3^2\right)},
\label{appendix_1_011}
\end{equation}
and permits one to recover eq~\ref{appendix_1_004}.

\subsubsection{Recursive construction of the integrity bases}
Computational
algorithms already exist to compute integrity bases\cite{b123,doc_03221,b165},
but they are limited to the case where the final representation is totally
symmetric. Furthermore, they are not very efficient for large dimensions.
 In contrast, the algorithm of section~\ref{recursion_ib} that parallels the recursive construction used for the generating functions 
can be applied  to compute efficiently the corresponding integrity basis.  Eq~\ref{appendix_1_010} contains more information than eq~\ref{appendix_1_003},
because it allows one to track the origin and the multiplicity of the different
terms. For example, the term $\left(1-t_1^2\right)$ in the
denominator of eq~\ref{appendix_1_010} comes from $\frac{1}{1-t_1^2}$ or
$\frac{t_1}{1-t_1^2}$.  Table~\ref{integrity_bases_egf} associates the $t_1²$
term in the denominator of these two fractions with the polynomial $x_1²$. As
a consequence, $x_1^2$ belongs to the denominator polynomials of
eq~\ref{appendix_1_010}.  The term $t_1t_2$ on the numerator of
eq~\ref{appendix_1_010} suggests a product of one numerator of degree one in
$x_1$ and one numerator of degree one in $x_2$ and leads to the conclusion
that the term $x_1x_2$ belongs to the numerator polynomials of
eq~\ref{appendix_1_010}. The integrity bases for our case example determined
with this method are given in Table~\ref{integrity_bases_toy_problem}.
Remembering that denominator polynomials can be multipled between themselves without
any restriction but that numerator polynomials only appear linearly, the lists
of invariant and $A_2$--covariant polynomials of degree $k$ in
Table~\ref{higher_degree_polynomials} are straightforwardly computed from the
integrity bases in Table~\ref{integrity_bases_toy_problem}. The data in
Table~\ref{integrity_bases_toy_problem} is enough to compute quickly a basis
for the vector space of invariant or $A_2$--covariant polynomials of any
degree. For example, the degree $5$, $A_2$--covariant  $x_2^2x_3^3=  (x_2^2x_3^2)\cdot x_3$,
is the product of a single numerator $A_2$--covariant,  $x_3$, with the product of denominator 
invariants $x_2^2x_3^2$.

\begin{table}[h!]
\begin{center}
\caption{Integrity bases for the two generating functions
  $M^{\mathrm{C_i}}\left(\Gamma^\mathrm{final};A_2 \oplus A_2 \oplus
  A_2;t\right)$ involved in the case example. 
\label{integrity_bases_toy_problem}}
\begin{tabular}{llll}
\hline \hline
$\Gamma^\mathrm{final}$ & Generating function & Denominator polynomials & Numerator polynomials \\
\hline
$A_1$ & $\frac{1+t_1t_2+t_1t_3+t_2t_3}{\left(1-t_1^2\right)\left(1-t_2^2\right)\left(1-t_3^2\right)}$ &
$\left\{x_1^2,x_2^2,x_3^2\right\}$ & $\left\{1,x_1x_2,x_1x_3,x_2x_3\right\}$
\\
$A_2$ & $\frac{t_1+t_2+t_3+t_1t_2t_3}{\left(1-t_1^2\right)\left(1-t_2^2\right)\left(1-t_3^2\right)}$ &
$\left\{x_1^2,x_2^2,x_3^2\right\}$ & $\left\{x_1,x_2,x_3,x_1x_2x_3\right\}$ \\
\hline \hline
\end{tabular}
\end{center}
\end{table}

\section{Application to the construction of integrity bases for \methane{} molecules}
Our main goal is to generate in the most economical way integrity bases for
representations of symmetry groups on vector spaces spanned by molecular
internal degrees of freedom. We focus, from now on, on the example of
\methane{} molecules, but the following method holds in general.  We consider
coordinates for the internal degrees of freedom adapted to the $\Td$ symmetry
point group of the molecule, which is isomorphous to the permutation group
$\mathcal{S}_4$.  For example, they can be the usual $\Td$--adapted
coordinates used in many studies on \methane{} molecules,\cite{doc_03239}
denoted by $S_1$, $S_{2a}$, $S_{2b}$, $S_{3x}$, $S_{3y}$, $S_{3z}$, $S_{4x}$,
$S_{4y}$, and $S_{4z}$.  $S_1$ transforms as the irrep $A_1$, the pair
$S_{2a}$, $S_{2b}$ transforms as $E$, while both triplets $S_{3x}$, $S_{3y}$,
$S_{3z}$ and $S_{4x}$, $S_{4y}$, and $S_{4z}$ transform as $F_2$.  So, the
representation of $\Td$ on the vector space $\Gamma^\mathrm{initial}:={\mathbb
  R}<S_{1},S_{2a},...,S_{4z}>$ generated by $S_{1},S_{2a},...,S_{4z}$ over the
field of real numbers (to which we restrict ourselves from now on, in view of
the applications) splits into a direct sum of irreps:
\begin{eqnarray}
\Gamma^\mathrm{initial}
&=&
{\mathbb R}<S_{1}>\oplus {\mathbb R}<S_{2a},S_{2b}>\oplus
{\mathbb R}<S_{3x},S_{3y},S_{3z}> \oplus {\mathbb R}<S_{4x}, S_{4y},S_{4z}>,
\nonumber \\
&=&
\Gamma^\mathrm{initial}_1 \oplus \Gamma^\mathrm{initial}_2 \oplus
\Gamma^\mathrm{initial}_3 \oplus \Gamma^\mathrm{initial}_4.
\label{9D-rep}
\end{eqnarray}

%Let us write ${\mathcal P}^{\Gamma^\mathrm{final}}$ for the vector space of such $\Gamma^\mathrm{final}$-covariants.
An extra \rm coordinate $S_5$ has to be added to map bi--univoquely the whole
nuclear configuration manifold, if the  coordinates are $\mathrm{O(3)}$--invariant
(such as linear combinations of bond distances and bond angles, and no
dihedral angle).\cite{a1776,doc_03705} In this case,  polynomials involved in the
computation of the PES, the DMS and other physically relevant quantities have
to be expressed as $P=P_0+P_1S_5+P_2S_5^2+P_3S_5^3$, where the $P_i$ are
polynomials in the coordinates 
$S_1,S_{2a},S_{2b},S_{3x},S_{3y},S_{3z},S_{4x},S_{4y},S_{4z}$. 

However, since $S_5$ can be chosen to carry the $A_1$ representation, this
extra--coordinate can be handled independently of the computation of
$\Gamma^\mathrm{final}$--covariants.  The same holds true for  $S_1$.
This allows us to reduce the problem to the study of ${\mathcal
  P}^{\Gamma^\mathrm{final}}$, where ${\mathcal P}$ is the polynomial algebra
generated by $S_{2a},S_{2b},S_{3x},S_{3y},S_{3z},S_{4x},S_{4y},S_{4z}$. Note
however, that the tabulated integrity bases provided as supplemental
material, Appendices A and B, as well as eq~\ref{molien-ch4-F2-num}
correspond to the full $9$--dimensional representation
$\Gamma^\mathrm{initial}$.

The octahedral group $\mathrm{O}$ and the group $\Td$ both belong to the
category of cubic point groups and share similar properties. Integrity bases
related to the Molien generating functions
$M\left(\Gamma^\mathrm{final};\Gamma^\mathrm{initial}_i;t\right)$, 
where $\Gamma^\mathrm{initial}_i$ and $\Gamma^\mathrm{final}$ are irreps,
are known for $\mathrm{O}$, see ref~\cite{a1392} and
Appendix~\ref{GF_IB_TD}. The denominator and numerator polynomials of these
integrity bases are the building blocks of the construction of the
integrity basis for the initial $8$--dimensional reducible representation,
$\Gamma^{\mathrm{initial}}_0:={\mathbb R}<S_{2a},S_{2b}>\oplus {\mathbb
  R}<S_{3x},S_{3y},S_{3z}>\oplus {\mathbb R}<S_{4x},
S_{4y},S_{4z}>=\Gamma^\mathrm{initial}_2 \oplus \Gamma^\mathrm{initial}_3
\oplus \Gamma^\mathrm{initial}_4$ of the tetrahedral group $\Td$.

\subsection{Denominator polynomials of the integrity bases}

Denominator polynomials of the integrity basis of a reducible representation is just the union
of the denominator polynomials of its irreducible subrepresentations.
The form of the $8$ denominator polynomials $f_2,...,f_9$ (the shift in the
indexing is motivated by the convention $f_1:=S_1$) for
$\Gamma^{\mathrm{initial}}_0$ is familiar.\cite{a1776} They consist in two denominator
polynomials of the module of $\Td$--invariant polynomials in $S_{2a},S_{2b}$,
${\mathbb R}[S_{2a},S_{2b}]^{\Td}$, three denominator
polynomials of ${\mathbb R}[S_{3x},S_{3y},S_{3z}]^{\Td}$ and of three
denominator polynomials of ${\mathbb R}[S_{4x},S_{4y},S_{4z}]^{\Td}$.
 We list them below by degrees of increasing order:

\begin{enumerate}
\item Degree 2:
\begin{equation}f_2:=\frac{
    S_{2a}^2+S_{2b}^2}{\sqrt{2}} \end{equation}
\begin{equation}f_3:=\frac{S_{3x}^2+S_{3y}^2+S_{3z}^2}{\sqrt{3}}\end{equation}
\begin{equation}f_4:=\frac{S_{4x}^2+S_{4y}^2+S_{4z}^2}{\sqrt{3}}\end{equation}
\item Degree 3:
\begin{equation}f_5:=\frac{-S_{2a}^3+3S_{2b}^2S_{2a}}{2}\end{equation}
\begin{equation}f_6:=S_{3x}S_{3y}S_{3z}\end{equation}
\begin{equation} f_7:=S_{4x}S_{4y}S_{4z}\end{equation}
\item Degree 4:
\begin{equation}f_8:=\frac{S_{3x}^4 + S_{3y}^4 + S_{3z}^4}{\sqrt{3}}\end{equation}
\begin{equation}f_9:=\frac{S_{4x}^4 + S_{4y}^4 + S_{4z}^4}{\sqrt{3}}\ .\end{equation}
\end{enumerate}

\subsection{Numerator polynomials of the integrity bases}

The Molien series for the action of $\Td$ on $\Gamma^{\mathrm{initial}}_0$ can
be directly computed using Burnside's generalization\cite{b167} of the
Molien's results.\cite{a1743} However, as suggested by the case example with
$\mathrm{C_i}$ symmetry, it is computationally more efficient to use
eq~\ref{molienmixed} to recursively construct the Molien generating functions
and the integrity bases. Setting $G=\Td$, a non--zero
$c_{\Gamma_\alpha,\Gamma_\beta}^{\Gamma}$ coefficient in the sum of
eq~\ref{molienmixed} relates at each step of the recursive algorithm to a
possible non--zero numerator $\Gamma_f$--covariant in the integrity basis of
the generating function
$M^{\Td}\left(\Gamma^\mathrm{final};\Gamma^\mathrm{initial}_1 \oplus
\Gamma^\mathrm{initial}_2 \oplus \Gamma^\mathrm{initial}_3 \oplus
\Gamma^\mathrm{initial}_4;t_1,t_2,t_3,t_4\right)$. The corresponding
polynomial is built by coupling previously obtained polynomials with
Clebsch--Gordan coefficients of the group $\Td$. \cite{doc_03458}

As an example, let us compute $M^{\Td}\left(E;F_2\oplus F_2;t_3,t_4\right)$.
The product table of the irreps of the group $\Td$ is given in
Table~\ref{product_table}. We can construct objects that transform according
to any of the five irreps from objects that carry the $F_2$ irrep.  As a
consequence, the five $M^{\Td}\left(\Gamma_\alpha;F_2;t\right)$, with
$\Gamma_\alpha$ an irrep, are non--zero.  Table~\ref{product_table} indicates
that the direct product $\Gamma_\alpha \times \Gamma_\beta$ contains the $E$
representation if and only if the pair
$\left(\Gamma_\alpha,\Gamma_\beta\right)$ belongs to the following set:
\begin{equation}
\left\{
\left(A_1,E\right),\,
\left(E,A_1\right),\,
\left(A_2,E\right),\,
\left(E,A_2\right),\,
\left(E,E\right),\,
\left(F_1,F_1\right),\,
\left(F_1,F_2\right),\,
\left(F_2,F_1\right),\,
\left(F_2,F_2\right)
\right\}.
\label{list_of_pairs}
\end{equation}

\begin{table}[h!]
\begin{center}
\caption{Product table of the irreps of the group $\Td$.
\label{product_table}}
\begin{tabular}{c|ccccc}
\hline \hline
 & $A_1$ & $A_2$ & $E$ & $F_1$ & $F_2$ \\
\hline
$A_1$ & $A_1$ & $A_2$ & $E$ & $F_1$ & $F_2$ \\
$A_2$ & $A_2$ & $A_1$ & $E$ & $F_2$ & $F_1$ \\
$E$ & $E$ & $E$ & $A_1 \oplus A_2 \oplus E$ & $F_1 \oplus F_2$ & $F_1 \oplus
F_2$ \\
$F_1$ & $F_1$ & $F_2$ & $F_1 \oplus F_2$ & $A_1 \oplus E \oplus F_1 \oplus
F_2$ & $A_2 \oplus E \oplus F_1 \oplus F_2$ \\
$F_2$ & $F_2$ & $F_1$ & $F_1 \oplus F_2$ & $A_2 \oplus E \oplus F_1 \oplus
F_2$ & $A_1 \oplus E \oplus F_1 \oplus F_2$ \\
\hline \hline
\end{tabular}
\end{center}
\end{table}

According to eq~\ref{molienmixed}, each of the nine pairs
$\left(\Gamma_\alpha,\Gamma_\beta\right)$ of eq~\ref{list_of_pairs}
contributes to a term
%$M^{\Td}\left(\Gamma_\alpha;F_2;t_3\right)M^{\Td}\left(\Gamma_\beta;F_2;t_4\right)$
in the expansion
\begin{equation}
M^{\Td}\left(E;F_2\oplus F_2;t_3,t_4\right)
=
\sum_{\Gamma_\alpha,\Gamma_\beta} c_{\Gamma_\alpha,\Gamma_\beta}^E
M^{\Td}\left(\Gamma_\alpha;F_2;t_3\right)M^{\Td}\left(\Gamma_\beta;F_2;t_4\right).
\label{exemple_Td}
\end{equation}
The expressions of the elementary generating functions
$M^{\Td}\left(\Gamma;F_2;t\right)$
are given in ref~\cite{a1392} and Appendix~\ref{GF_IB_TD}.
As an example, the pair $\left(F_2,F_1\right)$ of eq~\ref{list_of_pairs} will
give the following contribution in eq~\ref{exemple_Td}:
\begin{equation}
c_{F_2,F_1}^E M^{\Td}\left(F_2;F_2;t_3\right)M^{\Td}\left(F_1;F_2;t_4\right)=
\frac{\left(t_3+t^2_3+t^3_3\right)\left(t^3_4+t^4_4+t^5_4\right)}
{
\left(1-t_3^2\right)\left(1-t_3^3\right)\left(1-t_3^4\right)
\left(1-t_4^2\right)\left(1-t_4^3\right)\left(1-t_4^4\right)
}.
\label{exemple_F2_F1_E}
\end{equation}
The interpretation of the right--hand side of eq~\ref{exemple_F2_F1_E} in
terms of integrity basis suggests that the pair $\left(F_2,F_1\right)$ of
eq~\ref{list_of_pairs} will contribute to $6$ denominators and $2\times 9=18$
numerator polynomials to the integrity basis of $M^{\Td}\left(E;F_2\oplus
F_2;t_3,t_4\right)$. The $6$ denominator polynomials are simply the union of
the set of the denominator invariants associated to the
$M^{\Td}\left(F_2;F_2;t_3\right)$ and $M^{\Td}\left(F_1;F_2;t_4\right)$
elementary generating functions. Each product $t_3^{n_3}t_4^{n_4}$ in the
numerator of eq~\ref{exemple_F2_F1_E} corresponds to a numerator polynomial of
symmetry $E,i$ obtained by coupling via the Clebsch--Gordan coefficients of
the $\Td$ group the numerator polynomial of symmetry $F_2,j$, degree $n_3$
that belongs to the integrity basis of $M^{\Td}\left(F_2;F_2;t_3\right)$ with
the numerator polynomial of symmetry $F_1,k$, degree $n_4$ that belongs to the
integrity basis of $M^{\Td}\left(F_1;F_2;t_4\right)$. The expansion of the
product $\left(t_3+t^2_3+t^3_3\right)\left(t^3_4+t^4_4+t^5_4\right)$ contains
$9$ terms and each term contributes to two polynomials, one of symmetry type
$E,a$ and the other of symmetry type $E,b$, hence the $2\times9=18$
numerator polynomials.

This recursive algorithm has the advantage that only the integrity bases for
initial irreps, see Appendix~\ref{GF_IB_TD}, and the Clebsch--Gordan
coefficients of the group $\Td$ are required. In practice we couple first the
two symmetrized $F_2$ coordinates $S_{3x}$, $S_{3y}$, $S_{3z}$ and $S_{4x}$, $
S_{4y}$, $S_{4z}$. We then couple the results with the coordinates $S_{2a}$
and $S_{2b}$.  The fully coupled generating function for the $F_2$ final irrep
reads:
\begin{equation}
M^{\Td}\left(F_2;\Gamma^\mathrm{initial}_0;t\right)
=\frac{\mathcal{N}\left(F_2;\Gamma^\mathrm{initial}_0;t\right)}{(1-t^2)^3(1-t^3)^3(1-t^4)^2},
\label{molien-ch4-F2-8D}
\end{equation} 
with
\begin{eqnarray}
\lefteqn{\mathcal{N}\left(F_2;\Gamma^\mathrm{initial}_0;t\right)}
\nonumber \\
&=& 2t + 5t^2 + 12t^3 + 23t^4 + 41t^5 + 60t^6+  71t^7 + 71t^8 
\nonumber \\
&& + 60t^9 + 45t^{10} + 27t^{11} + 12t^{12} + 3t^{13}. 
\label{molien-ch4-F2-num}
\end{eqnarray}
Finally, to deal with the coordinate $S_1$, it suffices to note that
\begin{equation}
M^{\Td}\left(F_2;\Gamma^\mathrm{initial};t\right)
=\frac{M^{\Td}\left(F_2;\Gamma^\mathrm{initial}_0;t\right)}{(1-t)}.
\label{molien-ch4-F2-9D}
\end{equation}
The Molien series numerator coefficients for all irreps are given in
Table~\ref{generator_number}.

\begin{table}[h!]
\begin{center}
\caption{Numbers $n_k^{\Gamma^\mathrm{final}}$ of $\Gamma^\mathrm{final}$--covariant numerator
  polynomials of degree $k$ and dimensions
  $\dim\mathcal{P}^{\Gamma^\mathrm{final},i}_k$, $1\leq i \leq \left[ \Gamma^\mathrm{final}\right]$, of the
  vector spaces $\mathcal{P}^{\Gamma^\mathrm{final},i}_k$ of covariant polynomials of type
  $\Gamma^\mathrm{final},i$ and of degree $k$,
  $\Gamma^\mathrm{final}\in\{A_1,A_2,E,F_1,F_2\}$.  The total number $\sum_{k=0}^{15}
  n_k^{\Gamma^\mathrm{final}}$ of $\Gamma^\mathrm{final}$--covariant numerator polynomials is equal to
  $\left[ \Gamma^\mathrm{final}\right] \times \Pi_j d_j /|G| $, where $\left[ \Gamma^\mathrm{final}
    \right]$ is the dimension of the irrep $\Gamma^\mathrm{final}$,
  $|G|=24$ is the order of the group $\Td$, and $\Pi_j d_j=3456$ is the
  product of the degrees of the nine denominator polynomials. This result is a
  generalized version of proposition 2.3.6 of ref~\cite{b123}. It suffices to
  multiply the left--hand side of Eq.~(2.3.4) by the complex conjugate of the
  character of $\pi$ and to notice that this equals to $\left[\Gamma^\mathrm{final}\right]$
  for $\pi=Id$, see also Proposition 4.9 of ref~\cite{a1027}.
\label{generator_number}}

\begin{tabular}{c|cc|cc|cc|cc|cc}
\hline \hline
$\Gamma^\mathrm{final}$: &\multicolumn{2}{c|}{$A_1$} &\multicolumn{2}{c|}{$A_2$}&\multicolumn{2}{c|}{$E$}&\multicolumn{2}{c|}{$F_1$}&\multicolumn{2}{c}{$F_2$}\\
Degree $k$ & $n_k^{A_1}$ & $\dim\mathcal{P}^{A_1}_k$ & $n_k^{A_2}$ & $\dim\mathcal{P}^{A_2}_k$ & $n_k^{E}$
& $\dim\mathcal{P}^{E,i}_k$ & $n_k^{F_1}$ & $\dim\mathcal{P}^{F_1,i}_k$ & $n_k^{F_2}$ & $\dim\mathcal{P}^{F_2,i}_k$\\
\hline 
   0 &    1 & 1     & 0 & 0     & 0 & 0     &  0 & 0    &  0& 0          \\
   1 &    0 & 1     & 0 & 0     & 1 & 1     & 0  & 0    &  2& 2          \\
   2 &    1 & 5     & 0 & 0     & 4 & 5     & 3  & 3    &  5& 7          \\
   3 &    5 & 13    & 4 & 4     & 6 & 14    & 12 & 15   & 12& 25         \\
   4 &    9 & 33    & 8 & 12    &16 & 45    & 27 & 51   & 23& 69         \\
   5 &    12& 72    &15 & 39    &28 & 111   & 45 & 141  & 41& 177        \\
   6 &    18& 162   &26 & 101   &39 & 257   & 60 & 342  & 60& 400        \\
   7 &    21& 319   &24 & 226   &50 & 545   & 71 & 752  & 71& 848        \\
   8 &    24& 620   &21 & 470   &50 & 1090  & 71 & 1528 & 71& 1672       \\
   9 &    26& 1132  &18 & 918   &39 & 2040  & 60 & 2920 & 60& 3140       \\
   10 &   15& 1998  &12 & 1680  &28 & 3678  & 41 & 5298 & 45& 5610       \\
   11&    8 & 3384  & 9 & 2946  &16 & 6330  & 23 & 9210 & 27& 9654       \\
   12&    4 & 5587  & 5 & 4973  & 6 & 10545 & 12 & 15418& 12& 16022      \\
   13&    0 & 8912  & 1 & 8098  & 4 & 17010 &  5 & 24998&  3& 25822      \\
   14&    0 & 13912 & 0 & 12818 & 1 & 26730 &  2 & 39388&  0& 40472      \\
   15&    0 & 21185 & 1 & 19771 & 0 & 40935 &  0 & 60536&  0& 61960      \\
$n>15$&   0 &   & 0 &   & 0 &   &  0 &    &  0&      0   \\
\hline
Total&    144 &$\infty$& 144 &$\infty$& 288 &$\infty$&  432 &$\infty$&
432&$\infty$ \\
\hline \hline
\end{tabular}
\end{center}
\end{table}

As far as the $F_2$ representation is concerned, Table~\ref{generator_number}
tells that there are 432 numerator
polynomials of symmetry type $F_2,z$: $\left\{g_1^{F_2,z},...,g_{432}^{F_2,z}\right\}$ of
which $2$ are of degree one, $5$ of degree two, $12$ of degree three, and so
on.  We finally obtain that an arbitrary polynomial of symmetry type $F_2,m$,
$m\in\{x,y,z\}$ in the algebra spanned by the $S_1,\ldots,S_{4z}$ coordinates
will identify with a unique linear combination of monomials: 
\begin{equation}
f_1^{j_1}f_2^{j_2}...f_9^{j_9}g_k^{F_2,m} \ \ \left(j_1,\ldots j_9\right)\in\mathbb{N}^9,\ 1\leq k\leq 432.
\end{equation}

The lists of numerator polynomials for all irreps are provided as supplemental
material.\cite{doc_03735} They have been derived in a few seconds of CPU
time on a laptop by using the MAPLE computer algebra system.\cite{Maple} The
knowledge of the polynomials in our integrity bases is sufficient to generate
all the polynomials up to any degree, only multiplications between denominator
polynomials and one numerator polynomial are necessary. The recipe is given in
Appendix~\ref{toolbox}.  The gain with respect to classical methods of group
theory already shows up at degree $4$: we only need the 9 basic invariants and
the 16 $A_1$--covariants (i.e. secondary invariants) up to degree $4$, to
generate all $33$ linearly independent invariants of degree $4$ for
representation $\Gamma^\mathrm{initial}$, see Table~\ref{generator_number} and
compare with ref~\cite{doc_03183} where only a $6$--dimensional representation
is considered (the $S_{3x},S_{3y},S_{3z}$ coordinates are left out). In fact,
an integrity basis of $6$ basic invariants and $3$ secondary invariants can
generate $11$ linearly independent $A_1$--invariants of degree 4, which will
span the same vector space as those tabulated in the last table of
ref~\cite{doc_03183}. Similar remarks apply to the covariants. The gain
becomes rapidly more spectacular as the degree increases. PES of order $10$
have already been calculated for methane.\cite{doc_03595,doc_03073} There are
$1998$ linearly independent invariants of degree $10$ for representation
$\Gamma^\mathrm{initial}$. They can be generated with only the $9$ basic
invariants and $132$ secondary invariants. Similarly, EDMS for methane of
order $6$ have already appeared in the literature.\cite{doc_03484,doc_03538}
The $9$ basic invariants and $143$ $F_2,z$-covariant numerator polynomials of
degree less or equal to $6$ (see Table~\ref{generator_number}) are enough to
generate the $400$ linearly independent polynomials required to span the
vector space of $F_2,z$-covariant polynomials of sixth degree.

\section{Conclusion}

Our recursive method for constructing invariants and covariants blends ideas from the theory of invariants and from techniques used in applications of group theory to physics and chemistry.
We have determined for the first time integrity bases of the
$\Gamma^\mathrm{final}$--covariants of the group $\Td$ acting on the $9$ (or possibly $10$) symmetrized internal
coordinates of a \methane{} molecule. They are composed of nine algebraically
independent denominator polynomials and a finite number of
$\Gamma^\mathrm{final}$--covariant numerator polynomials given in the
supplemental material.\cite{doc_03735}
% ($432$ for the $z$-component of $F_2$-covariants given in supplementary
% material \cite{doc_03735}, $42$ up to order $4$ given in Appendix). These
% polynomials can be used to express (the smooth part of) symmetry-adapted,
% global PES for XY$_4$-type of molecules.

We have taken advantage of symmetry--adapted internal coordinates spanning the
reducible representation $A_1\oplus E\oplus F_2\oplus F_2$ of $\Td$, 
(and used in many studies of methane PES or EDMS as recalled in introduction),
 to construct an integrity basis for each final representation $\Gamma^\mathrm{final}$.
Integrity basis sets are first determined for each single, possibly
degenerate, irrep of the group.  These integrity bases
are coupled successively in a second step by using the
Clebsch--Gordan coefficients of the group $\Td$. 

This strategy to derive the $\Gamma^\mathrm{final}$--covariants is general
since the $\Gamma^\mathrm{final}$--covariant polynomials admit a Hironaka
decomposition \cite{a1027} for any finite group $G$.  Any ``internal
coordinate system'' (coordinates for internal degrees of freedom) $(q_i)_i$,
or internal displacement coordinates $(q_i-q_i^0)_i$, with respect to a
molecular reference configuration $(q_i^0)_i$ totally invariant under $G$, can
be symmetrized to obtain symmetry-adapted coordinates. Polynomials in the
latter coordinates can in turn be used to represent PES and other functions of
nuclear geometries. This is straightforward when such a function is
independent of the orientational coordinates (e.g. Euler angles) of the moving axes, 
like the PES, or the EDMS in the body-frame when using $\mathrm{O(3)}$-invariant 
coordinates: the moving--frame--dependent part being then all included in the 
direction--cosines which relate the EDM in the body-frame to the EDM in the
laboratory-frame.
However, for pentatomic and beyond the use of $\mathrm{O(3)}$-invariant coordinates 
necessarily implies auxiliary coordinates (such as $S_5$ in the case treated here) 
to cover the full configuration space \cite{b161} and one may wish to employ
moving-frame-dependent symmetry coordinates instead. Then, our approach can be
useful to obtain covariant bases of the permutation-invariant group, however,
the total symmetry group acting on the $3N-3$ (orientational + shape)
coordinates is only a semi-direct product of the permutation-inversion group
by SO(3) which makes the exploitation of symmetry for the EDMS in the
laboratory-frame more involved. The problem of body-frame definition and
singularities \cite{doc_03664} is out of the scope of the present paper.

So, in many cases of chemical interest, our approach makes available for the
study of global PES and other functions of nuclear geometric configurations
the recent tools of ring and invariant theory such as Cohen--Macaulay--type
properties and the effective computational tools of modern commutative
algebra,\cite{b164} which go far beyond the classical Molien series approach
in quantum chemistry.

Last but not least, our method based on integrity bases is more efficient than
classical methods of group theory based on the construction degree by degree
of the symmetry--adapted terms to be included in the potential energy surface
or the electric dipole moment surface.  All the required polynomials up to any
order can be generated by simple multiplications between polynomials in the
integrity bases of this paper in a direct manner as illustrated in
Appendix~\ref{toolbox}.

\section*{acknowledgements}
Financial support for the project \textit{Application de la Th\'eorie des
  Invariants \`a la Physique Mol\'eculaire} via a CNRS grant \textit{Projet
  Exploratoire Premier Soutien} (PEPS) \textit{Physique Th\'eorique et
  Interfaces} (PTI) is acknowledged. The first and third authors also acknowledge support from the grant CARMA ANR-12-BS01-0017.

\appendix

\section{Generating functions and corresponding integrity bases for
  irreducible representations of $\Td$\label{GF_IB_TD}}

The $\Td$ point group has five irreps: $A_1$, $A_2$,
$E$, $F_1$ and $F_2$. The irrep $E$ is doubly degenerate,
while the $F_1$ and $F_2$ irreps are triply degenerate.
The procedure detailed in section~\ref{ideas} is based on the knowledge of the generating functions
$M^{\Td}\left(\Gamma^\mathrm{final};\Gamma^\mathrm{initial};t\right)$,
where $\Gamma^\mathrm{initial}$ and $\Gamma^\mathrm{final}$ are irreps of the group
$\Td$. The coefficient $c_n$ in the Taylor expansion $c_0+c_1t+c_2t^2+\cdots$
of the generating function gives the number of linearly independent 
$\Gamma^\mathrm{final}$--covariant polynomials of degree $n$ that can be constructed from
the objects in the initial $\Gamma^\mathrm{initial}$ representation.

Each generating function $M^{\Td}\left(\Gamma^\mathrm{final};\Gamma^\mathrm{initial};t\right)$ is the
ratio of a numerator $\mathcal{N}\left(\Gamma^\mathrm{final};\Gamma^\mathrm{initial};t\right)$ over a
denominator $\mathcal{D}\left(\Gamma^\mathrm{initial};t\right)$:
\begin{equation}
M^{\Td}\left( \Gamma^\mathrm{final} ;\Gamma^\mathrm{initial} ; t \right)
=
\frac
{\mathcal{N}\left(\Gamma^\mathrm{final};\Gamma^\mathrm{initial};t\right)}
{\mathcal{D}\left(\Gamma^\mathrm{initial};t\right)}
=
\frac
{\sum\limits_{k=1}^{N} t^{\nu_k}}
{\prod\limits_{k=1}^{D} \left(1-t^{\delta_k}\right)}, 
\label{FG_def}
\end{equation}
with $\nu_k \in \mathbb{N}$ and $\delta_k \in \mathbb{N}\backslash{}\left\{0\right\}$.
The polynomial associated to a $\left(1-t^{\delta_k}\right)$ term in the
denominator is an invariant called a denominator polynomial of degree
$\delta_k$ and is noted $I^{\left(\delta_k\right)}\left(\Gamma^\mathrm{initial}\right)$.  The
polynomial associated to a $t^{\nu_k}$ term in the numerator is a
$\Gamma^\mathrm{final}$--covariant called a numerator polynomial of degree $\nu_k$ and is
noted $E^{(\nu_k)}\left(\Gamma^\mathrm{final};\Gamma^\mathrm{initial}\right)$ (when $\Gamma^\mathrm{final}$ is degenerate, 
$E^{(\nu_k)}\left(\Gamma^\mathrm{final};\Gamma^\mathrm{initial}\right)$ will be a vector gathering all the 
$\Gamma^\mathrm{final},i$--covariant numerator polynomials of degree $\nu_k$ for 
$i\in\{1,\ldots, \left[\Gamma^\mathrm{final}\right]\}$). According to the expression, eq~\ref{FG_def}, 
$D$ denominator polynomials and $N$ numerator polynomials are associated to the generating function, 
$M^{\Td}\left( \Gamma^\mathrm{final} ;\Gamma^\mathrm{initial} ; t \right)$.  

We will  closely follow the article of Patera, Sharp and Winternitz~\cite{a1392} for the notation for denominator and numerator polynomials, 
using  $\alpha, \beta, \gamma$  symbols for a chosen basis of each irrep.
However,  their table for octahedral tensors contains two errors for the degree eight
$E^{(8)}\left(\Gamma_4;\Gamma_4\right)$ and degree seven
$E^{(7)}\left(\Gamma_5;\Gamma_4\right)$ numerator polynomials.
With the definitions of polynomials given in ref~\cite{a1392}, the
following relation hold:
\begin{eqnarray}
E^{(8)}\left(\Gamma_4;\Gamma_4\right)_i
&=&
I^{(2)}\left(\Gamma_4\right)E^{(6)}\left(\Gamma_4;\Gamma_4\right)_i
\nonumber \\
&& -\frac{1}{2}I^{(2)}\left(\Gamma_4\right)^2E^{(4)}\left(\Gamma_4;\Gamma_4\right)_i
\nonumber \\
&& +\frac{1}{2}I^{(4)}\left(\Gamma_4\right)E^{(4)}\left(\Gamma_4;\Gamma_4\right)_i,
\label{ib_patera_1}
\end{eqnarray}
where the index $i$ stands either for $x$, $y$ or $z$.
The relation eq~\ref{ib_patera_1} indicates that the polynomial of degree eight
$E^{(8)}\left(\Gamma_4;\Gamma_4\right)$ has a decomposition in terms of
polynomials that are elements of the integrity basis associated to
$M^{\Td}\left(\Gamma_4;\Gamma_4;t\right)$. As a consequence, 
$E^{(8)}\left(\Gamma_4;\Gamma_4\right)$ does not enter the integrity basis.

The same is true for $E^{(7)}\left(\Gamma_5;\Gamma_4\right)$ and the integrity
basis associated to $M^{\Td}\left(\Gamma_5;\Gamma_4;t\right)$ due to following relation:

\begin{eqnarray}
E^{(7)}\left(\Gamma_5;\Gamma_4\right)_i
&=&
I^{(2)}\left(\Gamma_4\right)E^{(5)}\left(\Gamma_5;\Gamma_4\right)_i 
\nonumber \\
&& -\frac{1}{2}I^{(2)}\left(\Gamma_4\right)^2E^{(3)}\left(\Gamma_5;\Gamma_4\right)_i
\nonumber \\
&& +\frac{1}{2}I^{(4)}\left(\Gamma_4\right)E^{(3)}\left(\Gamma_5;\Gamma_4\right)_i.
\label{ib_patera_2}
\end{eqnarray}

A complete list of tables of both denominator and numerator polynomials for all the
initial $\Gamma^\mathrm{initial}$ and final $\Gamma^\mathrm{final}$ irreps is given
in the next sections.

\subsection{$\Gamma^\mathrm{initial}=A_1$ irreducible representation}

The denominator is $\mathcal{D}\left(A_1;t\right) = 1-t$.  The corresponding
denominator polynomial of degree one is $I^{\left(1\right)}\left(A_1\right) =
\alpha$.  The only non--zero numerator polynomial is
$\mathcal{N}\left(A_1;A_1;t\right)=1$.

\subsection{$\Gamma^\mathrm{initial}=A_2$ irreducible representation}

The denominator is $\mathcal{D}\left(A_2;t\right)=1-t^2$.
The corresponding denominator polynomial of degree two is
$I^{\left(2\right)}\left(A_2\right)=\alpha^2$.
Two numerator polynomials are non--zero: $\mathcal{N}\left(A_1;A_2;t\right)=1$
and $\mathcal{N}\left(A_2;A_2;t\right)=t$.
The $A_2$--covariant numerator polynomial of degree one is
$$
E^{\left(1\right)}\left(A_2;A_2\right)=\alpha.
$$

\subsection{$\Gamma^\mathrm{initial}=E$ irreducible representation}

The denominator is
$\mathcal{D}\left(E;t\right)=\left(1-t^2\right)\left(1-t^3\right)$.
The denominator polynomial of degree two is
$I^{\left(2\right)}\left(E\right)=\frac{\alpha^2+\beta^2}{\sqrt{2}}$ 
and the denominator polynomial of degree three is
$I^{\left(3\right)}\left(E\right)=\frac{-\alpha^3+3\alpha \beta^2}{2}$.
Three numerator polynomials are non--zero:
$\mathcal{N}\left(A_1;E;t\right)=1$,
$\mathcal{N}\left(A_2;E;t\right)=t^3$,
and
$\mathcal{N}\left(E;E;t\right)=t+t^2$.
The $A_2$--covariant numerator polynomial of degree three is
$$
E^{\left(3\right)}\left(A_2;E\right)=\frac{-3\alpha^2\beta+\beta^3}{2},
$$
and the two $E$--covariant numerator polynomials of degree one and two are
\begin{eqnarray}
E^{\left(1\right)}\left(E;E\right)
&=&
\left(\begin{array}{c}\alpha\\ \beta \end{array}\right),
\nonumber \\
E^{\left(2\right)}\left(E;E\right)
&=&
\frac{1}{\sqrt{2}}\left(\begin{array}{c}-\alpha^2+\beta^2\\ 2\alpha\beta \end{array}\right).
\nonumber
\end{eqnarray}

\subsection{$\Gamma^\mathrm{initial}=F_1$ irreducible representation}

The denominator is
$\mathcal{D}\left(F_1;t\right)=\left(1-t^2\right)\left(1-t^4\right)\left(1-t^6\right)$.
The denominator polynomial of degree two is
$I^{\left(2\right)}\left(F_1\right)=\frac{\alpha^2+\beta^2+\gamma^2}{\sqrt{3}}$,
the denominator polynomial of degree four is 
$I^{\left(4\right)}\left(F_1\right)=\frac{\alpha^4+\beta^4+\gamma^4}{\sqrt{3}}$ 
and the denominator polynomial of degree six is
$I^{\left(6\right)}\left(F_1\right)=\frac{\alpha^6+\beta^6+\gamma^6}{\sqrt{3}}$.
The numerator polynomials are
$\mathcal{N}\left(A_1;F_1;t\right)=1+t^9$,
$\mathcal{N}\left(A_2;F_1;t\right)=t^3+t^6$,
$\mathcal{N}\left(E;F_1;t\right)=t^2+t^4+t^5+t^7$,
$\mathcal{N}\left(F_1;F_1;t\right)=t+t^3+t^4+t^5+t^6+t^8$,
and
$\mathcal{N}\left(F_2;F_1;t\right)=t^2+t^3+t^4+t^5+t^6+t^7$.
The invariant numerator polynomial of degree nine is
$$
E^{\left(9\right)}\left(A_1;F_1\right)=\frac{1}{\sqrt{6}} \alpha \beta \gamma
\left(\alpha^2-\beta^2\right)
\left(\beta^2-\gamma^2\right)
\left(\gamma^2-\alpha^2\right),
$$
the two $A_2$--covariant numerator polynomials of degree three and six are
\begin{eqnarray}
E^{\left(3\right)}\left(A_2;F_1\right)&=&\alpha\beta\gamma,
\nonumber \\
E^{\left(6\right)}\left(A_2;F_1\right)&=&\frac{1}{\sqrt{6}}\left(\alpha^2-\beta^2\right),
\left(\beta^2-\gamma^2\right)
\left(\gamma^2-\alpha^2\right),
\nonumber
\end{eqnarray}
the four $E$--covariant numerator polynomials of degree two, four, five, and
seven are:
\begin{eqnarray}
E^{\left(2\right)}\left(E;F_1\right)&=&\frac{1}{\sqrt{6}}\left(\begin{array}{c}
\alpha^2+\beta^2-2\gamma^2
\\ \sqrt{3}\left(-\alpha^2+\beta^2\right) \end{array}\right),
\nonumber \\
E^{\left(4\right)}\left(E;F_1\right)&=&\frac{1}{\sqrt{6}}\left(\begin{array}{c}
\alpha^4+\beta^4-2\gamma^4
\\ \sqrt{3}\left(-\alpha^4+\beta^4\right)\end{array}\right),
\nonumber \\
E^{\left(5\right)}\left(E;F_1\right)&=&\frac{1}{\sqrt{6}}\alpha\beta\gamma\left(\begin{array}{c}
\sqrt{3}\left(\alpha^2-\beta^2\right)
\\ \alpha^2+\beta^2-2\gamma^2 \end{array}\right),
\nonumber \\
E^{\left(7\right)}\left(E;F_1\right)&=&\frac{1}{\sqrt{6}}\alpha\beta\gamma\left(\begin{array}{c}
\sqrt{3}\left(\alpha^4-\beta^4\right)
\\ \alpha^4+\beta^4-2\gamma^4\end{array}\right),
\nonumber
\end{eqnarray}
the six $F_1$--covariant numerator polynomials of degree one, three, four,
five, six, and eight are
\begin{eqnarray}
E^{\left(1\right)}\left(F_1;F_1\right)&=&\left(\begin{array}{c}
\alpha  \\ \beta \\ \gamma \end{array}\right),
\nonumber \\
E^{\left(3\right)}\left(F_1;F_1\right)&=&\left(\begin{array}{c}
\alpha^3  \\ \beta^3 \\ \gamma^3 \end{array}\right),
\nonumber \\
E^{\left(4\right)}\left(F_1;F_1\right)&=&\frac{1}{\sqrt{2}}\left(\begin{array}{c}
\left(\beta^2-\gamma^2\right)\beta\gamma  \\ 
\left(\gamma^2-\alpha^2\right)\gamma\alpha\\ 
\left(\alpha^2-\beta^2\right)\alpha\beta
\end{array}\right),
\nonumber \\
E^{\left(5\right)}\left(F_1;F_1\right)&=&\left(\begin{array}{c}
\alpha^5  \\ \beta^5\\ \gamma^5\end{array}\right),
\nonumber \\
E^{\left(6\right)}\left(F_1;F_1\right)&=&\frac{1}{\sqrt{2}}\left(\begin{array}{c}
\left(\beta^4-\gamma^4\right)\beta\gamma  \\ 
\left(\gamma^4-\alpha^4\right)\gamma\alpha\\ 
\left(\alpha^4-\beta^4\right)\alpha\beta
 \end{array}\right),
\nonumber \\
E^{\left(8\right)}\left(F_1;F_1\right)&=&\frac{1}{\sqrt{2}}
\alpha \beta \gamma
\left(\begin{array}{c}
\left(\beta^4-\gamma^4\right)\alpha  \\ 
\left(\gamma^4-\alpha^4\right)\beta\\ 
\left(\alpha^4-\beta^4\right)\gamma
\end{array}\right),
\nonumber
\end{eqnarray}
the six $F_2$--covariant numerator polynomials of degree two, three, four,
five, six, and seven are
\begin{eqnarray}
E^{\left(2\right)}\left(F_2;F_1\right)&=&\left(\begin{array}{c}
\beta \gamma  \\ \gamma \alpha\\ \alpha \beta \end{array}\right)
\nonumber \\
E^{\left(3\right)}\left(F_2;F_1\right)&=&\frac{1}{\sqrt{2}}\left(\begin{array}{c}
\left(\beta^2-\gamma^2\right)\alpha  \\ 
\left(\gamma^2-\alpha^2\right)\beta\\ 
\left(\alpha^2-\beta^2\right)\gamma
\end{array}\right),
\nonumber \\
E^{\left(4\right)}\left(F_2;F_1\right)&=&
\alpha \beta \gamma
\left(\begin{array}{c}
\alpha
\\ \beta\\\gamma \end{array}\right),
\nonumber \\
E^{\left(5\right)}\left(F_2;F_1\right)&=&\frac{1}{\sqrt{2}}\left(\begin{array}{c}
\left(\beta^4-\gamma^4\right)\alpha  \\ 
\left(\gamma^4-\alpha^4\right)\beta\\ 
\left(\alpha^4-\beta^4\right)\gamma
\end{array}\right),
\nonumber \\
E^{\left(6\right)}\left(F_2;F_1\right)&=&
\alpha \beta \gamma
\left(\begin{array}{c}
\alpha^3
\\ \beta^3\\\gamma^3 \end{array}\right),
\nonumber \\
E^{\left(7\right)}\left(F_2;F_1\right)&=&\frac{1}{\sqrt{2}}
\alpha\beta\gamma
\left(\begin{array}{c}
\left(\beta^2-\gamma^2\right)\beta\gamma  \\ 
\left(\gamma^2-\alpha^2\right)\alpha\gamma \\
\left(\alpha^2-\beta^2\right)\alpha\beta
\end{array}\right).
\nonumber
\end{eqnarray}

\subsection{$\Gamma^\mathrm{initial}=F_2$ irreducible representation}

The denominator is
$\mathcal{D}\left(F_2;t\right)=\left(1-t^2\right)\left(1-t^3\right)\left(1-t^4\right)$.
The denominator polynomial of degree two is
$I^{\left(2\right)}\left(F_2\right)=\frac{\alpha^2+\beta^2+\gamma^2}{\sqrt{3}}$, 
the denominator polynomial of degree three is
$I^{\left(3\right)}\left(F_2\right)=\alpha\beta\gamma$
and the denominator polynomial of degree four is
$I^{\left(4\right)}\left(F_2\right)=\frac{\alpha^4+\beta^4+\gamma^4}{\sqrt{3}}$.
The numerator polynomials are
$\mathcal{N}\left(A_1;F_2;t\right)=1$,
$\mathcal{N}\left(A_2;F_2;t\right)=t^6$,
$\mathcal{N}\left(E;F_2;t\right)=t^2+t^4$,
$\mathcal{N}\left(F_1;F_2;t\right)=t^3+t^4+t^5$,
and
$\mathcal{N}\left(F_2;F_2;t\right)=t+t^2+t^3$.
The $A_2$--covariant numerator polynomial of degree six is
$$
E^{\left(6\right)}\left(A_2;F_2\right)=
\frac{1}{\sqrt{6}}
\left(\alpha^2-\beta^2\right)
\left(\beta^2-\gamma^2\right)
\left(\gamma^2-\alpha^2\right),
$$
the two $E$--covariant numerator polynomials of degree two and four are
\begin{eqnarray}
E^{\left(2\right)}\left(E;F_2\right)
&=&
\frac{1}{\sqrt{6}}
\left(\begin{array}{c}
\alpha^2+\beta^2-2\gamma^2
\\ \sqrt{3}\left(-\alpha^2+\beta^2\right)
\end{array}\right),
\nonumber \\
E^{\left(4\right)}\left(E;F_2\right)
&=&
\frac{1}{\sqrt{6}}
\left(\begin{array}{c}
\alpha^4+\beta^4-2\gamma^4
\\ \sqrt{3}\left(-\alpha^4+\beta^4\right)
\end{array}\right),
\nonumber
\end{eqnarray}
the four $F_1$--covariant numerator polynomials of degree three, four and five are
\begin{eqnarray}
E^{\left(3\right)}\left(F_1;F_2\right)
&=&
\frac{1}{\sqrt{2}}
\left(\begin{array}{c}
\left(\beta^2-\gamma^2\right)\alpha  \\ 
\left(\gamma^2-\alpha^2\right)\beta\\ 
\left(\alpha^2-\beta^2\right)\gamma
\end{array}\right),
\nonumber \\
E^{\left(4\right)}\left(F_1;F_2\right)
&=&
\frac{1}{\sqrt{2}}
\left(\begin{array}{c}
\left(\beta^2-\gamma^2\right)\beta\gamma  \\ 
\left(\gamma^2-\alpha^2\right)\gamma\alpha\\ 
\left(\alpha^2-\beta^2\right)\alpha\beta
\end{array}\right),
\nonumber \\
E^{\left(5\right)}\left(F_1;F_2\right)
&=&
\frac{1}{\sqrt{2}}
\left(\begin{array}{c}
\left(\beta^2-\gamma^2\right)\alpha^3  \\ 
\left(\gamma^2-\alpha^2\right)\beta^3\\ 
\left(\alpha^2-\beta^2\right)\gamma^3
\end{array}\right),
\nonumber
\end{eqnarray}
the three $F_2$--covariant numerator polynomials of degree one, two, and three are
\begin{eqnarray}
E^{\left(1\right)}\left(F_2;F_2\right)
&=&
\left(\begin{array}{c}
\alpha \\ \beta \\  \gamma \end{array}\right),
\nonumber \\
E^{\left(2\right)}\left(F_2;F_2\right)
&=&
\left(\begin{array}{c}
\beta \gamma  \\ \gamma\alpha\\ \alpha\beta\end{array}\right),
\nonumber \\
E^{\left(3\right)}\left(F_2;F_2\right)
&=&
\left(\begin{array}{c}
\alpha^3 \\ \beta^3 \\  \gamma^3 \end{array}\right).
\nonumber
\end{eqnarray}

\section{Application of the integrity base for $F_2$--covariant polynomials: 
Representation of the  electric dipole moment surface of a tetrahedral \methane{} molecule
\label{toolbox}}

\subsection{Introduction}

Appendix~\ref{toolbox} gives an application of the integrity basis for
$F_2$--covariant polynomials of tetrahedral \methane{} molecules. The
integrity basis determined in this paper contains the denominator polynomials
$f_i$, $1\leq i\leq 9$, listed in the main text and the auxiliary numerators
published in the file \texttt{symmetries\_A1\_A2\_E\_F1\_F2.txt} available as
supplemental material.\cite{doc_03735} This example can be transposed to any
other final irrep $\Gamma^\mathrm{final}$.

The  electric dipole moment surface of a tetrahedral \methane{}
molecule can be built as a linear combination of $F_2$--covariant polynomials
of total degree less than $d_\mathrm{max}$ in the coordinates that span the
representation, $\Gamma^\mathrm{initial}$, of eq~\ref{9D-rep}. The integer $d_\mathrm{max}$ is the
order of the expansion.  The generating function for the number of
$F_2$--covariant polynomials built from this representation
reads (see eqs~\ref{molien-ch4-F2-8D} to~\ref{molien-ch4-F2-9D}):
$$
\frac{2t + 5t^2 + 12t^3 + 23t^4 + 41t^5 + 60t^6+  71t^7 + 71t^8 + 60t^9 +
  45t^{10} + 27t^{11} + 12t^{12} + 3t^{13}}
{\left(1-t\right)\left(1-t^2\right)^3\left(1-t^3\right)^3\left(1-t^4\right)^2},
$$
whose Taylor expansion up to order four is given by:
\begin{equation}
2t+7t^2+25t^3+69t^4+\cdots.
\label{taylor_expansion}
\end{equation}
The coefficients in eq~\ref{taylor_expansion} mean that there are $2$
(respectively $7$, $25$, and $69$) linearly independent $F_2,\alpha$--covariant
polynomials of degree one (respectively two, three, and four), $\alpha\in\{x,y,z\}$. 
We now detail the construction of these $103$ $F_2,x$ polynomials.
 The $F_2,y$ and $F_2,z$ polynomials may be built using the same procedure.

The expansion of the $F_2,x$-EDMS up to
order four is a linear combination of $103$ $F_2,x$--polynomials:
\begin{eqnarray}
\lefteqn{\mu_{F_2,x}\left(S_1,S_{2a},S_{2b},S_{3x},S_{3y},S_{3z},S_{4x},S_{4y},S_{4z}\right)}
\nonumber \\
&=&
\sum_{i=1}^{103} c_i^{F_2,x} \times
p_{i}^{F_2,x}\left(S_1,S_{2a},S_{2b},S_{3x},S_{3y},S_{3z},S_{4x},S_{4y},S_{4z}\right).
\label{expansion}
\end{eqnarray}
The coefficients $c_i^{F_2,x}$ of eq~\ref{expansion} are to be determined by
fitting the expression to either experimental or \textit{ab initio} data.
We know that the $103$, $F_2,x$--polynomials, $p_{i}^{F_2,x}$, can be written as a product of denominator polynomials powered to any
positive integer value, and a single numerator
polynomial. 
So, the polynomials that enter the expansion of the
$F_2,x$ component of the EDMS can all be taken of the form:
\begin{equation}
\varphi_{k,l}^{F_2,x} \times f_1^{n_1}f_2^{n_2}\cdots f_9^{n_9},
\label{expression}
\end{equation}
where the  $(\varphi_{k,l}^{F_2,x})_{1\leq l \leq
n_k^{F_2}}$ denotes the numerator polynomials of degree $k$, (we change the notation with respect to the main text to include explicitly the degree $k$). Their numbers, $n_k^{F_2}$, are given in the column labelled $F_2$ of Table~\ref{generator_number}.
%On the other side, a numerator can only appear linearly.
 Sets of linearly independent $p_{i}^{F_2,x}$ are listed below by degrees.  We recall that $f_1$ is a polynomial of degree one, $f_2$,
$f_3$, and $f_4$ are three polynomials of degree two, $f_5$,
$f_6$, and $f_7$ are three polynomials of degree three, and $f_8$, $f_9$ are
two polynomials of degree four.
 
\subsection{Degree one}
The $2$ $F_2,x$ linearly independent polynomials of total degree one compatible
with eq~\ref{expression} are $p_{1}^{F_2,x}=\varphi_{1,1}^{F_2,x}$ and
$p_{2}^{F_2,x}=\varphi_{1,2}^{F_2,x}$.

\subsection{Degree two}
The $7$ $F_2,x$ linearly independent polynomials of total degree two compatible
with eq~\ref{expression} are:
$$
\begin{array}{rclrclrclrclrcl}
p_{3}^{F_2,x}&=&\varphi_{2,1}^{F_2,x}, &
p_{4}^{F_2,x}&=&\varphi_{2,2}^{F_2,x}, &
p_{5}^{F_2,x}&=&\varphi_{2,3}^{F_2,x}, &
p_{6}^{F_2,x}&=&\varphi_{2,4}^{F_2,x}, &
p_{7}^{F_2,x}&=&\varphi_{2,5}^{F_2,x}, \\
p_{8}^{F_2,x}&=&\varphi_{1,1}^{F_2,x}f_1, &
p_{9}^{F_2,x}&=&\varphi_{1,2}^{F_2,x}f_1.
\end{array}
$$

\subsection{Degree three}
The $25$ $F_2,x$ linearly independent polynomials of total degree three compatible
with eq~\ref{expression} are:
$$
\begin{array}{rclrclrclrclrcl}
p_{10}^{F_2,x}&=&\varphi_{3,1}^{F_2,x}, &
p_{11}^{F_2,x}&=&\varphi_{3,2}^{F_2,x}, &
p_{12}^{F_2,x}&=&\varphi_{3,3}^{F_2,x}, &
p_{13}^{F_2,x}&=&\varphi_{3,4}^{F_2,x}, &
p_{14}^{F_2,x}&=&\varphi_{3,5}^{F_2,x}, \\
p_{15}^{F_2,x}&=&\varphi_{3,6}^{F_2,x}, &
p_{16}^{F_2,x}&=&\varphi_{3,7}^{F_2,x}, &
p_{17}^{F_2,x}&=&\varphi_{3,8}^{F_2,x}, &
p_{18}^{F_2,x}&=&\varphi_{3,9}^{F_2,x}, &
p_{19}^{F_2,x}&=&\varphi_{3,{10}}^{F_2,x}, \\
p_{20}^{F_2,x}&=&\varphi_{3,{11}}^{F_2,x}, &
p_{21}^{F_2,x}&=&\varphi_{3,{12}}^{F_2,x}, &
p_{22}^{F_2,x}&=&\varphi_{2,1}^{F_2,x} f_1, &
p_{23}^{F_2,x}&=&\varphi_{2,2}^{F_2,x} f_1, &
p_{24}^{F_2,x}&=&\varphi_{2,3}^{F_2,x} f_1, \\
p_{25}^{F_2,x}&=&\varphi_{2,4}^{F_2,x} f_1, &
p_{26}^{F_2,x}&=&\varphi_{2,5}^{F_2,x} f_1, &
p_{27}^{F_2,x}&=&\varphi_{1,1}^{F_2,x} f_1^2, &
p_{28}^{F_2,x}&=&\varphi_{1,2}^{F_2,x} f_1^2, &
p_{29}^{F_2,x}&=&\varphi_{1,1}^{F_2,x} f_2, \\
p_{30}^{F_2,x}&=&\varphi_{1,2}^{F_2,x} f_2, &
p_{31}^{F_2,x}&=&\varphi_{1,1}^{F_2,x} f_3, &
p_{32}^{F_2,x}&=&\varphi_{1,2}^{F_2,x} f_3, &
p_{33}^{F_2,x}&=&\varphi_{1,1}^{F_2,x} f_4, &
p_{34}^{F_2,x}&=&\varphi_{1,2}^{F_2,x} f_4.
\end{array}
$$

\subsection{Degree four}
The $69$ $F_2,x$ linearly independent polynomials of total degree four compatible
with eq~\ref{expression} are:
% taylor( (t*phi_1_1+t*phi_1_2 + t^2*phi_2_1 + t^2*phi_2_2 + t^2*phi_2_3 + t^2*phi_2_4 + t^2*phi_2_5 + t^3*phi_3_1 + t^3*phi_3_2 + t^3*phi_3_3 + t^3*phi_3_4 +t^3*phi_3_5+ t^3*phi_3_6 + t^3*phi_3_7 + t^3*phi_3_8 + t^3*phi_3_9 + t^3*phi_3_10 + t^3*phi_3_11 + t^3*phi_3_12 + t^4*phi_4_1 + t^4*phi_4_2 + t^4*phi_4_3 + t^4*phi_4_4 +t^4*phi_4_5+ t^4*phi_4_6 + t^4*phi_4_7 + t^4*phi_4_8 + t^4*phi_4_9 + t^4*phi_4_10 + t^4*phi_4_11 + t^4*phi_4_12 + t^4*phi_4_13 + t^4*phi_4_14 +t^4*phi_4_15+ t^4*phi_4_16 + t^4*phi_4_17 + t^4*phi_4_18 + t^4*phi_4_19 + t^4*phi_4_20 + t^4*phi_4_21 + t^4*phi_4_22 + t^4*phi_4_23)/( (1-t*f1)*(1-t^2*f2)*(1-t^2*f3)*(1-t^2*f4)*(1-t^3*f5)*(1-t^3*f6)*(1-t^3*f7)*(1-t^4*f8)*(1-t^4*f9)), t=0, 5);
\begin{small}$$
\begin{array}{rclrclrclrclrcl}
p_{35}^{F_2,x}&=&\varphi_{4,1}^{F_2,x}, &
p_{36}^{F_2,x}&=&\varphi_{4,2}^{F_2,x}, &
p_{37}^{F_2,x}&=&\varphi_{4,3}^{F_2,x}, &
p_{38}^{F_2,x}&=&\varphi_{4,4}^{F_2,x}, &
p_{39}^{F_2,x}&=&\varphi_{4,5}^{F_2,x}, \\
p_{40}^{F_2,x}&=&\varphi_{4,6}^{F_2,x}, &
p_{41}^{F_2,x}&=&\varphi_{4,7}^{F_2,x}, &
p_{42}^{F_2,x}&=&\varphi_{4,8}^{F_2,x}, &
p_{43}^{F_2,x}&=&\varphi_{4,9}^{F_2,x}, &
p_{44}^{F_2,x}&=&\varphi_{4,10}^{F_2,x}, \\
p_{45}^{F_2,x}&=&\varphi_{4,11}^{F_2,x}, &
p_{46}^{F_2,x}&=&\varphi_{4,12}^{F_2,x}, & 
p_{47}^{F_2,x}&=&\varphi_{4,13}^{F_2,x}, &
p_{48}^{F_2,x}&=&\varphi_{4,14}^{F_2,x}, &
p_{49}^{F_2,x}&=&\varphi_{4,15}^{F_2,x}, \\
p_{50}^{F_2,x}&=&\varphi_{4,16}^{F_2,x}, &
p_{51}^{F_2,x}&=&\varphi_{4,17}^{F_2,x}, &
p_{52}^{F_2,x}&=&\varphi_{4,18}^{F_2,x}, &
p_{53}^{F_2,x}&=&\varphi_{4,19}^{F_2,x}, &
p_{54}^{F_2,x}&=&\varphi_{4,20}^{F_2,x}, \\
p_{55}^{F_2,x}&=&\varphi_{4,21}^{F_2,x}, &
p_{56}^{F_2,x}&=&\varphi_{4,22}^{F_2,x}, &
p_{57}^{F_2,x}&=&\varphi_{4,23}^{F_2,x}, &
p_{58}^{F_2,x}&=&\varphi_{2,1}^{F_2,x} f_2, &
p_{59}^{F_2,x}&=&\varphi_{2,2}^{F_2,x} f_2, \\
p_{60}^{F_2,x}&=&\varphi_{2,3}^{F_2,x} f_2, &
p_{61}^{F_2,x}&=&\varphi_{2,4}^{F_2,x} f_2, &
p_{62}^{F_2,x}&=&\varphi_{2,5}^{F_2,x} f_2, &
p_{63}^{F_2,x}&=&\varphi_{2,1}^{F_2,x} f_3, &
p_{64}^{F_2,x}&=&\varphi_{2,2}^{F_2,x} f_3, \\
p_{65}^{F_2,x}&=&\varphi_{2,3}^{F_2,x} f_3, &
p_{66}^{F_2,x}&=&\varphi_{2,4}^{F_2,x} f_3, &
p_{67}^{F_2,x}&=&\varphi_{2,5}^{F_2,x} f_3, &
p_{68}^{F_2,x}&=&\varphi_{2,1}^{F_2,x} f_5, &
p_{69}^{F_2,x}&=&\varphi_{2,2}^{F_2,x} f_5, \\
p_{70}^{F_2,x}&=&\varphi_{2,3}^{F_2,x} f_5, &
p_{71}^{F_2,x}&=&\varphi_{2,4}^{F_2,x} f_5, &
p_{72}^{F_2,x}&=&\varphi_{2,5}^{F_2,x} f_5, &
p_{73}^{F_2,x}&=&\varphi_{1,1}^{F_2,x} f_5, &
p_{74}^{F_2,x}&=&\varphi_{1,2}^{F_2,x} f_5, \\
p_{75}^{F_2,x}&=&\varphi_{1,1}^{F_2,x} f_6, &
p_{76}^{F_2,x}&=&\varphi_{1,2}^{F_2,x} f_6, &
p_{77}^{F_2,x}&=&\varphi_{1,1}^{F_2,x} f_7, &
p_{78}^{F_2,x}&=&\varphi_{1,2}^{F_2,x} f_7, &
p_{79}^{F_2,x}&=&\varphi_{1,1}^{F_2,x} f_1 f_2, \\
p_{80}^{F_2,x}&=&\varphi_{1,2}^{F_2,x} f_1 f_2, &
p_{81}^{F_2,x}&=&\varphi_{1,1}^{F_2,x} f_1 f_3, &
p_{82}^{F_2,x}&=&\varphi_{1,2}^{F_2,x} f_1 f_3, &
p_{83}^{F_2,x}&=&\varphi_{1,1}^{F_2,x} f_1 f_4, &
p_{84}^{F_2,x}&=&\varphi_{1,2}^{F_2,x} f_1 f_4, \\
p_{85}^{F_2,x}&=&\varphi_{3,1}^{F_2,x} f_1, &
p_{86}^{F_2,x}&=&\varphi_{3,2}^{F_2,x} f_1, &
p_{87}^{F_2,x}&=&\varphi_{3,3}^{F_2,x} f_1, &
p_{88}^{F_2,x}&=&\varphi_{3,4}^{F_2,x} f_1, &
p_{89}^{F_2,x}&=&\varphi_{3,5}^{F_2,x} f_1, \\
p_{90}^{F_2,x}&=&\varphi_{3,6}^{F_2,x} f_1, &
p_{91}^{F_2,x}&=&\varphi_{3,7}^{F_2,x} f_1, &
p_{92}^{F_2,x}&=&\varphi_{3,8}^{F_2,x} f_1, &
p_{93}^{F_2,x}&=&\varphi_{3,9}^{F_2,x} f_1, &
p_{94}^{F_2,x}&=&\varphi_{3,10}^{F_2,x} f_1, \\
p_{95}^{F_2,x}&=&\varphi_{3,11}^{F_2,x} f_1, &
p_{96}^{F_2,x}&=&\varphi_{3,12}^{F_2,x} f_1, &
p_{97}^{F_2,x}&=&\varphi_{2,1}^{F_2,x} f_1^2, &
p_{98}^{F_2,x}&=&\varphi_{2,2}^{F_2,x} f_1^2, &
p_{99}^{F_2,x}&=&\varphi_{2,3}^{F_2,x} f_1^2, \\
p_{100}^{F_2,x}&=&\varphi_{2,4}^{F_2,x} f_1^2, &
p_{101}^{F_2,x}&=&\varphi_{2,5}^{F_2,x} f_1^2, &
p_{102}^{F_2,x}&=&\varphi_{1,1}^{F_2,x} f_1^3, &
p_{103}^{F_2,x}&=&\varphi_{1,2}^{F_2,x} f_1^3. 
\end{array}
$$  \end{small}

\end{document}